\documentclass[aps,prd,preprint,nofootinbib]{revtex4-2}

\usepackage{graphicx}
\usepackage[dvipsnames]{xcolor}
\usepackage{lineno,hyperref}
\usepackage{amssymb}
\usepackage{amsmath}
\usepackage{xspace}
\usepackage{upgreek}

\begin{document}
\title{Impact-parameter-dependent solutions to the Balitsky-Kovchegov equation at next-to-leading order}
\author{J. Cepila}
\author{J. G. Contreras}
\author{M. Matas}
\author{M. Vaculciak}
\email{matej.vaculciak@fjfi.cvut.cz}
\affiliation{Czech Technical University in Prague}

\begin{abstract}
{A stable numerical solution of the impact-parameter-dependent next-to-leading order  Balitsky-Kovchegov equation is presented for the first time. The rapidity evolution of the dipole amplitude is discussed in detail. Dipole amplitude properties, such as the evolution speed or anomalous dimension behaviour, are studied as a function of the impact parameter and the dipole size and compared to solutions of the impact-parameter-dependent leading-order Balitsky-Kovchegov equation with the collinearly improved kernel. The next-to-leading evolution presented here also strongly suppresses the Coulomb tails compared to the collinearly improved and leading order solutions. The origin of this behaviour is explored and potential avenues to improve the next-to-leading order kernel are pointed out.}
\end{abstract}


\maketitle

\section{Introduction}
Precise deep-inelastic scattering data from electron--proton collisions at the HERA accelerator show that the gluon content of the proton grows as a power law for decreasing values of Bjorken-$x$~\cite{H1:2009pze}. This behaviour cannot continue unchecked to arbitrarily small Bjorken-$x$, or high energy, correspondingly, as it would violate basic properties of quantum chromodynamics (QCD). It was predicted long ago that the growth of the gluon distribution in hadrons would be tamed by the recombination of gluons~\cite{Gribov:1983ivg,Mueller:1985wy}. This phenomenon is called \textit{saturation} and its experimental confirmation is an important research topic in theoretical studies of perturbative QCD (see e.g. Ref.~\cite{Morreale:2021pnn} for a recent review) as well as in the experimental arena. Here the LHC has recently provided several interesting measurements at high energies~\cite{Contreras:2015dqa,Klein:2019qfb,ALICE:2022wpn}, and the community is eagerly awaiting the start of operations of the Electron--Ion Collider, which is currently in construction and features the search for gluon saturation as one of its pillars~\cite{Accardi:2012qut,AbdulKhalek:2021gbh}. 

An appropriate framework to study the high-energy limit of QCD is the dipole picture~\cite{Kovchegov:2012mbw} where a quark-antiquark pair forms a colour dipole that interacts with the colour fields of the probed target. In the case of photon-induced interactions, such a dipole is provided by a fluctuation of a photon into a quark-antiquark state. In the eikonal approximation, the propagation of the dipole through the target can be described in the Colour Glass Condensate approach, see e.g. Ref.~\cite{Gelis:2010nm}, giving rise to the B-JIMWLK equations~\cite{Balitsky:1995ub,Jalilian-Marian:1997qno,Jalilian-Marian:1997jhx,Jalilian-Marian:1997ubg,Kovner:2000pt,Weigert:2000gi,Iancu:2000hn,Iancu:2001ad,Ferreiro:2001qy}. In the limit of a large number of colours, these equations reduce to the Balitsky-Kovchegov (BK) equation~\cite{Balitsky:1995ub,Kovchegov:1999yj,Kovchegov:1999ua}, which gives the evolution of the dipole amplitude $N(Y, \vec{r},\vec{b})$ in $Y$, where $Y$ denotes the rapidity of the dipole (projectile). The amplitude itself then describes the scattering of the incoming colour dipole of transverse size $\vec{r}$ with a hadronic target located at a distance given by the impact parameter vector $\vec{b}$.

The original BK equation was computed at the leading order (LO). Later on, the running of the coupling constant was included in the BK formalism~\cite{Albacete:2007sm,Albacete:2007yr} and the resulting equation was then used to successfully describe data from deep-inelastic scattering~\cite{Albacete:2009fh,Albacete:2010sy}. In this last work, and most of those following it, the dipole amplitude was taken as depending only on rapidity and the size of the dipole $r\equiv |\vec{r}|$, that is $N(Y, r)$, representing the interaction of the dipole with an infinite homogeneous target. Other higher-order effects were computed by resumming the radiative corrections enhanced by large double transverse logarithms and including single-logarithmic corrections~\cite{Iancu:2015vea,Iancu:2015joa}; the resulting collinearly improved (CI) BK equation was again used to obtain a good description of deep-inelastic scattering data~\cite{Iancu:2015joa}. 

Two ingredients are needed to contrast the BK equation with data at next-to-leading order (NLO) precision: the computation of impact factors at NLO and the NLO BK equation. Regarding the former, there has been considerable advance in the last few years in computing NLO impact factors for different observables, see e.g. Refs.~\cite{Beuf:2016wdz,Beuf:2017bpd,Beuf:2021qqa,Beuf:2021srj,Beuf:2022ndu,Mantysaari:2022kdm,Beuf:2024msh}. As for the latter, the NLO BK equation has been originally obtained in Ref.~\cite{Balitsky:2007feb}. Solutions of the form $N(Y, r)$ were studied numerically in Ref.~\cite{Lappi:2015fma}, where it was found that this equation produced unphysical results. The problem was identified as the large double transverse logarithmic NLO corrections, which are present in the full NLO BK equation with a negative sign. A viable form of the NLO BK equation was presented in Ref.~\cite{Lappi:2016fmu}, where the $N(Y, r)$ solutions were demonstrated to be stable when including the resummation of single and double logarithms in a procedure akin to what was done at LO~\cite{Iancu:2015vea,Iancu:2015joa}. 
 
 The first attempt to include the dependence on the impact parameter in the dipole amplitude, interpreted physically as a target of finite size and with a profile in impact parameter, was presented in Ref.~\cite{Golec-Biernat:2003naj} where the original LO BK equation was studied. It was found that in this case the amplitude develops unhealthy Coulomb tails, a power-law growth of the amplitude with rapidity at large impact parameters. This behaviour was not physical and precluded the use of these solutions for phenomenology. A modified BK equation introducing a cut-off in the kernel as a proxy for kinematic constraints~\cite{Motyka:2009gi} was studied~\cite{Berger:2010sh} and used to describe deep-inelastic scattering data~\cite{Berger:2011ew} as well as diffractive exclusive vector meson photoproduction~\cite{Berger:2012wx}. However, the issue of unphysical Coulomb tails remained. A reasonable description of measurements was achieved only through the use of this cut-off and the addition of a non-perturbative contribution. With the aforementioned advent of the collinear corrections to the BK equation, our group studied impact-parameter dependent solutions of the form $N(Y,r,b)$, with $b \equiv |\vec{b}|$, to the CI BK equation, finding that the Coulomb tails are strongly suppressed in this case~\cite{Cepila:2018faq}. This allows for a good description of data without any extra ad hoc ingredients~\cite{Cepila:2018faq,Bendova:2019psy}. The suppression of the Coulomb tails is explained by the fact that the CI kernel suppresses large daughter dipoles due to the time ordering imposed by the resummation. A similar suppression of power-law tails was found in a restricted range---that covers the region where measurement data exist---of the evolution in Ref.~\cite{Contreras:2019vox}.
 Recently, we extended this work to solutions of the form $N(Y,r,b,\theta)$ with $\theta$ representing the dipole orientation with respect to the impact parameter~\cite{Cepila:2023pvh}. For those studies, we used a variation of the BK equation with the evolution performed not in the rapidity of the projectile, but on the rapidity of the target~\cite{Ducloue:2019ezk}.

In this article, we extend the results from Ref.~\cite{Lappi:2016fmu} to the numerical study of solutions to the NLO BK equation by including the dependence on the size of the impact parameter in the dipole amplitude, i.e. we explore solutions of NLO BK of the form $N(Y, r,b)$. 
We find that the solutions are numerically stable. These NLO solutions have a slower evolution speed than the CI solutions. They exhibit a more stable anomalous dimension as rapidity increases than in the CI case. The NLO solutions also show a suppression of the power-law Coulomb tails, making them a viable ingredient for future comparison of observables with data. 

The article is organised as follows: in Sec.~\ref{sec:bk} the BK equation at NLO and its leading order CI variant are briefly introduced, Sec.~\ref{sec:solving} introduces the method to obtain the stable numerical solutions of the BK equation, and Sec.~\ref{sec:sol} presents the obtained dipole amplitudes and discusses their properties. 

\section{The BK equation \label{sec:bk}}
\subsection{The BK equation at next-to-leading-order}
\begin{figure}
    \includegraphics[width=.98\linewidth]{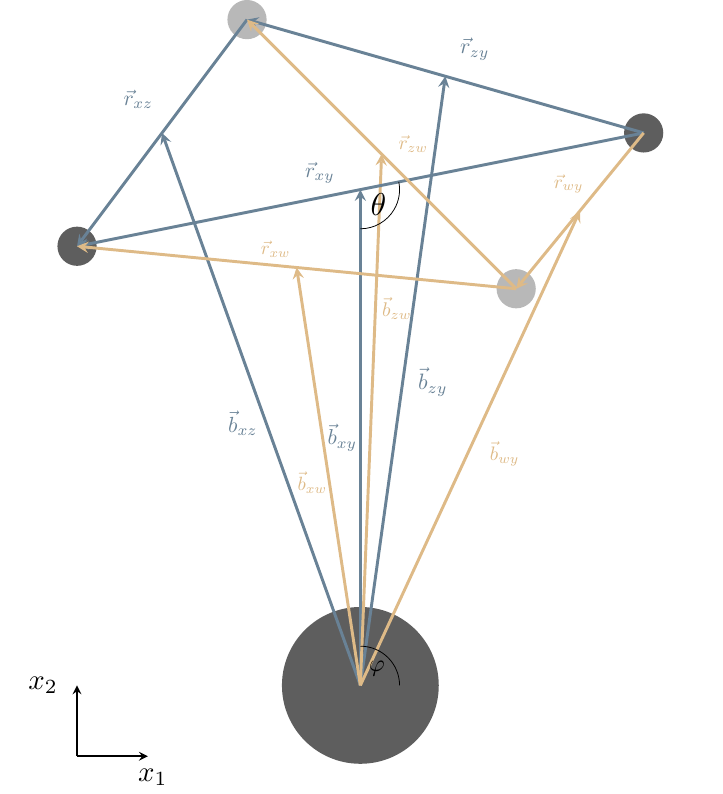}
    \caption{\label{fig:layout}The transversal structure of the mother and daughter dipoles within the next-to-leading evolution. The leading order contribution is highlighted in blue, while the additional gluonic emission and the associated dipoles are in orange. The angle $\theta$ controls the tilting of the mother dipole with respect to the target, and $\varphi$ is the angle of the impact parameter. In this work, we assume an isotropic target and neglect the $\theta$ dependence, so the scattering amplitude is a function of rapidity $Y$, dipole size $r$, and impact parameter $b$ only.}
\end{figure}

Including an additional gluon contribution in the BK evolution, the projectile-rapidity next-to-leading order BK equation was derived in~\cite{Balitsky:2007feb} and can be written as
\begin{eqnarray}
    \partial_Y N_{xy} &=& \int d^2 z K_0 \Big[ N_{xz} + N_{zy} - N_{xy} - N_{xz}N_{zy} \Big] 
    \nonumber \\
    &+& \int d^2 z d^2 w K_1 \Big[N_{wy} + N_{zw} - N_{zy} - N_{xz}N_{zw} - N_{xz}N_{wy} 
    \nonumber \\
    &&\hspace{75pt} - N_{zw}N_{wy} + N_{xz}N_{zy} + N_{xz}N_{zw}N_{wy} \Big]
    \nonumber \\
    &+& \int d^2 z d^2 w K_2 \Big[ N_{xw} - N_{xz} - N_{zy}N_{xw} + N_{xz}N_{zy} \Big],
    \label{eq:nlo-bk}
\end{eqnarray}
where $N_{a} \equiv N(Y, r_{a}, b_{a}, \theta_{a}, \varphi_{a})$. Here $a$ keeps track of the coordinates at the dipole ends, e.g. $xy$. Throughout this work, we assume an isotropic target, lifting the dependence of the scattering amplitude on the angle $\varphi_a$. Furthermore, we integrate out the angular orientation of the dipole with respect to the target $\theta_a$, leading to an amplitude that depends on the dipole size and impact parameter only $N_{a} \equiv N(Y, r_{a}, b_{a})$.
An example of such a coordinate set-up is shown in Fig.~\ref{fig:layout}. Throughout the presented calculations, both for the NLO and CI cases, we fix the orientation of the parent dipole to $\theta = \pi/2$.

Resumming contributions from large transverse logarithms~\cite{Iancu:2015joa} and double logarithmic terms~\cite{Iancu:2015vea} the kernels take the form
\begin{equation}
    K_0 = K_\mathrm{rc} K_\mathrm{STL} K_\mathrm{DLA} + K_\mathrm{sub} + K_\mathrm{fin},
\end{equation}
where the LO running-coupling kernel is 
\begin{widetext}
\begin{equation}
    K_\mathrm{rc} = \frac{\alpha_\mathrm{s}(r_{xy}) N_\mathrm{C}}{2 \pi^2} 
    \Big[
    \frac{r_{xy}^2}{r_{xz}^2 r_{zy}^2} 
    + 
    \frac{1}{r_{xz}^2} \Big(\frac{\alpha_\mathrm{s}(r_{xz})}{\alpha_\mathrm{s}(r_{zy})} - 1\Big) 
    +
    \frac{1}{r_{zy}^2} \Big(\frac{\alpha_\mathrm{s}(r_{zy})}{\alpha_\mathrm{s}(r_{xz})} - 1\Big)
    \Big], 
\end{equation}
\end{widetext}
with $N_\mathrm{C}$ representing the number of colours. The part of the kernel present within the collinearly improved form of the BK equation is 
\begin{eqnarray}
    K_\mathrm{DLA} &= \begin{cases}
        \frac{\mathrm{J}_1(2\sqrt{\bar{\alpha}_\mathrm{s} \rho^2})}{\sqrt{\bar{\alpha}_\mathrm{s} \rho^2}} 
        & \rho^2 > 0 \\ 
        \frac{\mathrm{I}_1(2\sqrt{-\bar{\alpha}_\mathrm{s} \rho^2})}{\sqrt{- \bar{\alpha}_\mathrm{s} \rho^2}} 
        & \rho^2 < 0,
    \end{cases}
\end{eqnarray}
where $\rho^2 = \ln \frac{r_{xz}^2}{r_{xy}^2} \ln \frac{r_{zy}^2}{r_{xy}^2}$. The remaining terms unique to the NLO evolution with a single-quark emission are
\begin{eqnarray}
    K_\mathrm{STL} &=& \exp \Big[-\bar{\alpha}_\mathrm{s} A_1 \Big|\ln \Big(\frac{C_\mathrm{sub} r_{xy}^2}{\min (r_{xz}^2, r_{zy}^2)} \Big)\Big| \Big],
    \nonumber \\ 
    K_\mathrm{sub} &=& \frac{\bar{\alpha}_\mathrm{s}}{2 \pi}
    \Big[
    \bar{\alpha}_\mathrm{s} A_1 \Big|\ln \Big(\frac{C_\mathrm{sub} r_{xy}^2}{\min (r_{xz}^2, r_{zy}^2)} \Big)\Big|
    \Big]
    \frac{r_{xy}^2}{r_{xz}^2 r_{zy}^2},
    \nonumber \\ 
    K_\mathrm{fin} &=& \frac{\bar{\alpha}_\mathrm{s}^2}{8 \pi} \frac{r_{xy}^2}{r_{xz}^2 r_{zy}^2}
    \Big(
    \frac{67}{9} - \frac{\pi^2}{3} - \frac{10n_\mathrm{f}}{9N_\mathrm{C}}
    \Big),
\end{eqnarray}
where $A_1 = \frac{11}{12}$, $C_\mathrm{sub} = 0.65$, $n_\mathrm{f}$ represents the number of active flavours and $\bar{\alpha}_\mathrm{s} = \frac{N_\mathrm{C}}{\pi} \alpha (\min(r_{xy}, r_{xz}, r_{zy}))$. The parts of the kernel responsible for a two-gluon emission represented by a double integral over $dz$ and $dw$ in the evolution are
\begin{eqnarray}
    K_1 = \frac{\alpha_\mathrm{s}^2(r_{xy}) N_\mathrm{C}^2}{8 \pi^4} 
            \Big[
                &-&\frac{2}{r_{zw}^4} \Big(
                    \frac{r_{xz}^2 r_{wy}^2 + r_{xw}^2 r_{zy}^2 - 4 r_{xy}^2 r_{zw}^2}{r_{zw}^4 (r_{xz}^2 r_{wy}^2 - r_{xw}^2 r_{zy}^2)} + \nonumber \\
                &+&
                    \frac{r_{xy}^4}{r_{xz}^2 r_{wy}^2 (r_{xz}^2 r_{wy}^2 - r_{xw}^2 r_{zy}^2)}
                    +
                    \frac{r_{xy}^2}{r_{xz}^2 r_{wy}^2 r_{zw}^2}
                 \Big)
              \ln \frac{r_{xz}^2 r_{wy}^2}{r_{xw}^2 r_{zy}^2}
            \Big],
\end{eqnarray}
and
\begin{equation}
    K_2 = \frac{n_\mathrm{f} N_\mathrm{C} \alpha_\mathrm{s}^2(r_{xy})}{8 \pi^4} 
            \Big[
                \frac{2}{r_{zw}^4} -
                \Big(
                    \frac{r_{xw}^2 r_{zy}^2 + r_{wy}^2 r_{xz}^2 - r_{xy}^2 r_{zw}^2}{r_{zw}^4 (r_{xz}^2 r_{wy}^2 - r_{xw}^2 r_{zy}^2)}
                 \Big)
              \ln \frac{r_{xz}^2 r_{wy}^2}{r_{xw}^2 r_{zy}^2}
            \Big],
\end{equation}
as described in more detail in Ref.~\cite{Lappi:2016fmu}.

The expression we use for the running coupling throughout this work is 
\begin{equation}
\alpha_{\mathrm{s}, n_\mathrm{f}} (r^{2}) = \frac{4\pi}{\beta_{0, n_\mathrm{f}}\ln\left(\frac{4C^{2}}{r^{2}\Lambda ^{2}_{n_\mathrm{f}}}\right)},\label{eq:alph}
\end{equation}
where $C^{2}$, the so-called infrared regulator, is usually fit to data, and we fix it as $C = 2.5$. Here,
\begin{equation}\label{eq:beta}
\beta_{0, n_\mathrm{f}} = 11 - \frac{2}{3}n_\mathrm{f}.
\end{equation} 
$\Lambda^{2}_{n_\mathrm{f}}$ is determined from the number of active flavours as~\cite{Albacete:2010sy}
\begin{equation}
\Lambda_{n_\mathrm{f}-1}=(m_\mathrm{f})^{1-\frac{\beta_{0,n_\mathrm{f}}}{\beta_{0,n_\mathrm{f}-1}}}(\Lambda_{n_\mathrm{f}})^{\frac{\beta_{0,n_\mathrm{f}}}{\beta_{0,n_\mathrm{f}-1}}},
\end{equation}
where the active flavours are given by the dipole size and the quark mass $m_\mathrm{f}$ as 
\begin{equation}\label{eq:active}
r^{2} < \frac{4C^{2}}{m_\mathrm{f}^{2}}.
\end{equation}

We consider up to 5 active flavours, $N_\mathrm{C} = 3$, and set $m_\mathrm{f}=0.1$\,GeV$/c^2$ for the light quarks, and \mbox{$m_\mathrm{c}=1.3$\,GeV$/c^2$}, $m_\mathrm{b}=4.5$\,GeV$/c^2$ for the massive quarks. To fix the value of  $\Lambda_{5}$, we use the measured value of $\alpha_\mathrm{s}(M_\mathrm{Z})=0.1189 \pm 0.0017$ at the mass of the Z$^{0}$ boson \mbox{$M_\mathrm{Z}=91.18$\,GeV$/c^2$~\cite{ParticleDataGroup:2014cgo}}. 
From a certain dipole size, namely when
\begin{equation}
    r^2 \geq \frac{4C^2}{\Lambda_3^2} \exp{\bigg[- \frac{4\pi}{\beta_{0,3}\alpha_{s,0}}\bigg]},
\end{equation}
the growth of the coupling constant needs to be frozen at a fixed value $\alpha_{\mathrm{s},0}$~\cite{Quiroga-Arias:2011dfd,Albacete:2010sy}. In the scope of this work, the choice is $\alpha_{\mathrm{s},0} = 1$.

The initial condition we chose for the evolution combines the Gaussian shape of the proton profile with the GBW dependence for the behaviour of the scattering amplitude as
\begin{equation}\label{eq:ciinitialeq}
N(r, b,Y=0) = 1 - \exp\left(-\frac{1}{2}\frac{Q^2_{\mathrm{s},0}}{4}r^2 T(b_{\mathrm{q}_1},b_{\mathrm{q}_2})\right),
\end{equation}
where $b_{\mathrm{q}_i}$ are the impact parameters of the quark and anti-quark forming the dipole and
\begin{equation}
T(b_{\mathrm{q}_1},b_{\mathrm{q}_2})= \left[\exp\left(-\frac{b_{\mathrm{q}_1}^2}{2B_\mathrm{G}}\right) + \exp\left(-\frac{b_{\mathrm{q}_2}^2}{2B_\mathrm{G}}\right)\right].
\end{equation}
Two free parameters are required here: the saturation scale at $Y=0$, $Q^2_{\mathrm{s},0}$ and $B_\mathrm{G}$, defining the size of the proton~\cite{Bendova:2019psy}. We set them as $Q^2_{\mathrm{s},0} = 0.38$\,GeV$^2$ and $B_\mathrm{G} = 3.8$\,GeV$^{-2}$.

\subsection{The CI BK equation}
The CI BK equation, whose solutions are shown below to compare with the NLO solutions, reads 
\begin{equation}
    \partial_Y N_{xy} = \int d^2 z K_\mathrm{CI} \Big[ N_{xz} + N_{zy} - N_{xy} - N_{xz}N_{zy} \Big],
    \label{eq:ci-bk}
\end{equation}
and the collinearly improved kernel is
\begin{equation}
    K_\mathrm{CI} = \frac{\bar{\alpha}_\mathrm{s}}{2 \pi} \frac{r_{xy}^2}{r_{xz}^2 r_{zy}^2} 
    \bigg( \frac{r_{xy}^2}{\min (r_{xz}^2, r_{zy}^2)}\bigg)^{\pm \bar{\alpha}_\mathrm{s} A_1} 
    \frac{J_1(2\sqrt{\bar{\alpha}_\mathrm{s} |\rho^2|})}{\sqrt{\bar{\alpha}_\mathrm{s} |\rho^2|}}.
\end{equation}
Here, the exponent is positive if $r_{xy}^2 < \min(r_{xz}^2, r_{zy}^2)$, otherwise the negative sign is used. 

\section{Solving the equation\label{sec:solving}}
The numerical approach to solving the NLO BK equation is based on Euler's method (the Runge-Kutta method of the first order). This is because the 4th-order Runge-Kutta method is not applicable for the next-to-leading BK equation in the form used throughout our previous work (see~\cite{Cepila:2015qea, Cepila:2023pvh}) and the numerical precision is not affected by this change.

In the past works, solving the BK equation considering only the dependence on the dipole size, it was advantageous to align the grid of the integration variable $\vec{r}_{xz}$ with the grid sampling the scattering amplitude $N(r_{xy})$. During the evolution calculation, the values $N(r_{xy})$, $N(r_{xz})$ and $N(r_{zy})$ are needed and having these grids aligned, one only needs to interpolate the latter, gaining a calculation speed-up. 

However, this advantage drops out when including the impact-parameter dependence, as one rarely comes across a combination of the dipole size $\vec{r}_{xz}$ and impact parameter $\vec{b}_{xz}$ such that both coincide with the pre-calculated grid. It is therefore better to split the grids for sampling and integration altogether. This offers a significant improvement in the numerical calculation since one can tune the grids independently, to optimise for both precision and calculation speed at the cost of keeping track of the additional parameters.

Throughout this work, a logarithmic grid is used to sample the dipole amplitude in dipole size $r$ and impact parameter $b$, while in rapidity $Y$ it is sampled linearly. 
For the integration grids, polar coordinates are used, linear in angle and logarithmic in radius.

Although the dipole amplitude is in principle restricted to $N(Y, r, b) \in [0,1]$, rounding errors and small imprecisions can cause an under- or overflow of these bounds during the calculation. Since the origin of these errors is purely numerical, such values are set to 0 or 1, correspondingly.

Several terms in the denominator of the NLO kernel are divergent when $r_{xz}^2 r_{wy}^2 = r_{xw}^2 r_{zy}^2$, presenting a potential numerical problem. This typically leads to the introduction of ad hoc cut-off parameters to manually tame the large contributions. We were able to avoid this altogether and accept arbitrarily large kernel values throughout the calculation. The only exception is the unlikely event of hitting one of the aforementioned divergent points; in this case, the contribution of this point is naturally set to 0.

This level of numerical stability was achieved by introducing a relative offset of all the grids sampling the transverse plane in Fig.~\ref{fig:layout}. The integration polar grids are displaced as shown in Fig.~\ref{fig:grids} to avoid symmetrical configurations (for more details see Appendix~\ref{app:solving}).

\section{\label{sec:sol}Solutions to the BK equation at NLO}
\subsection{\label{sec:NYrb}Evolution of the dipole scattering amplitude}
\begin{figure*}[!ht]
    \includegraphics[width=.49\linewidth]{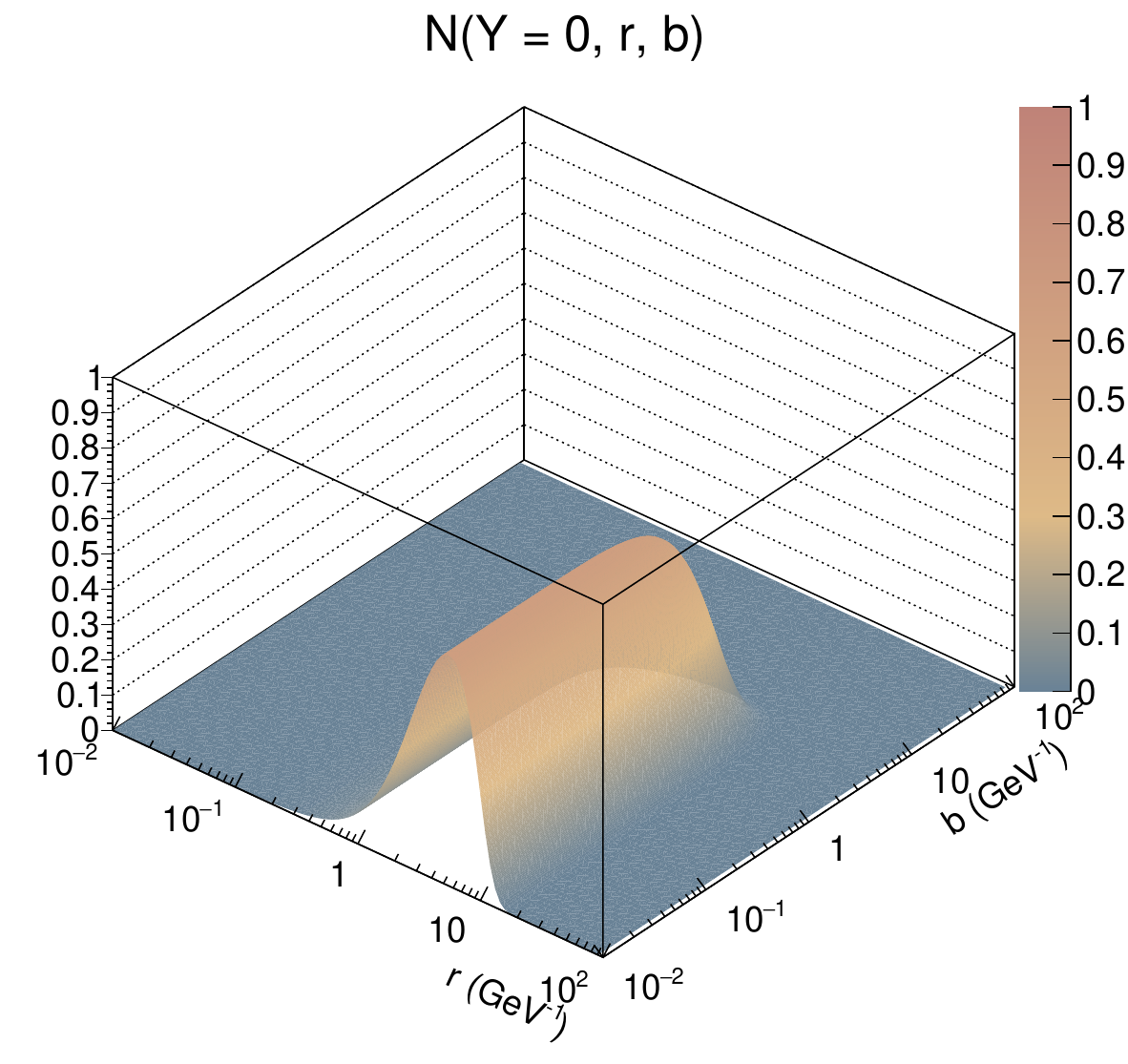}
    \includegraphics[width=.49\linewidth]{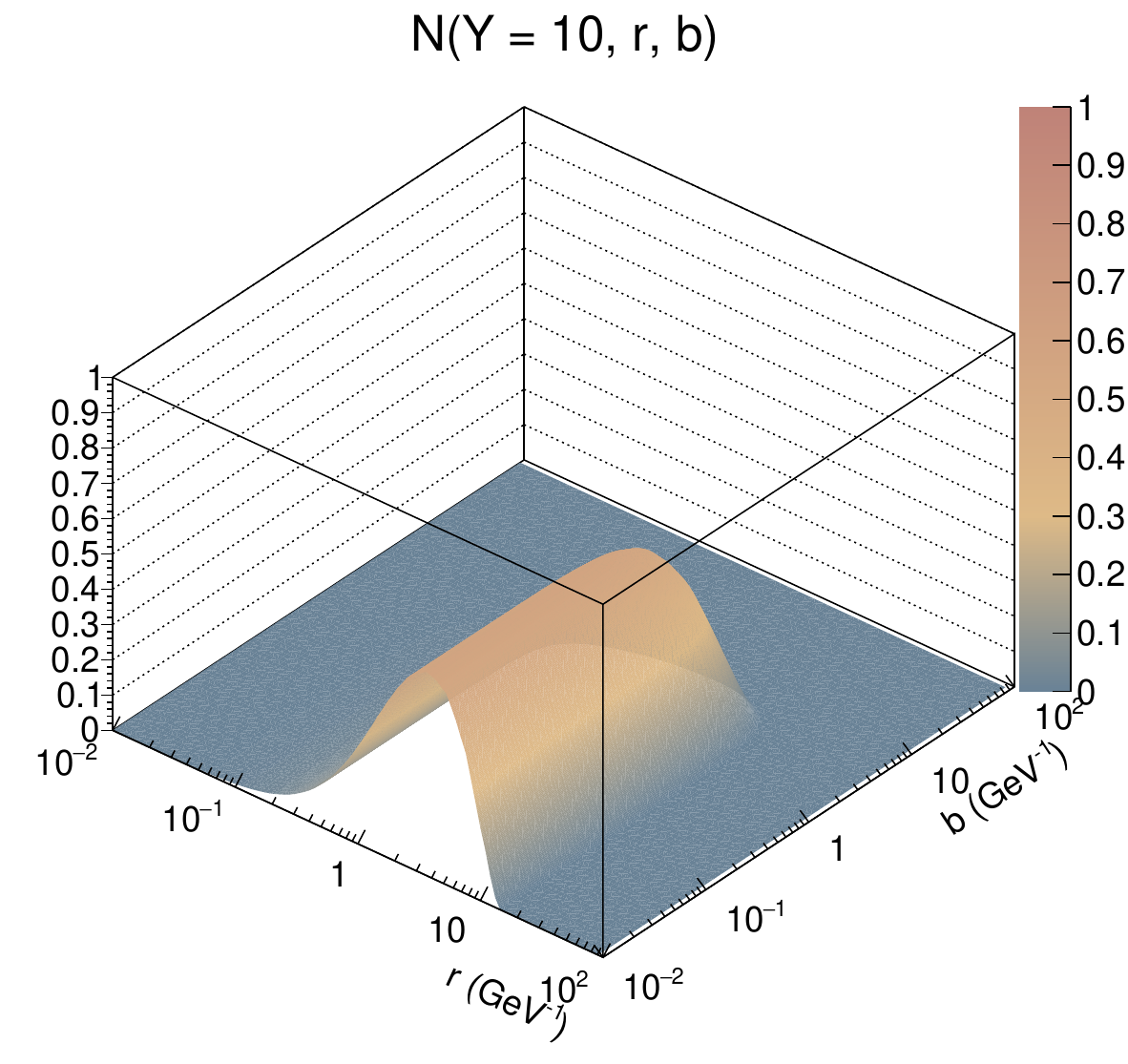}\\
    \includegraphics[width=.49\linewidth]{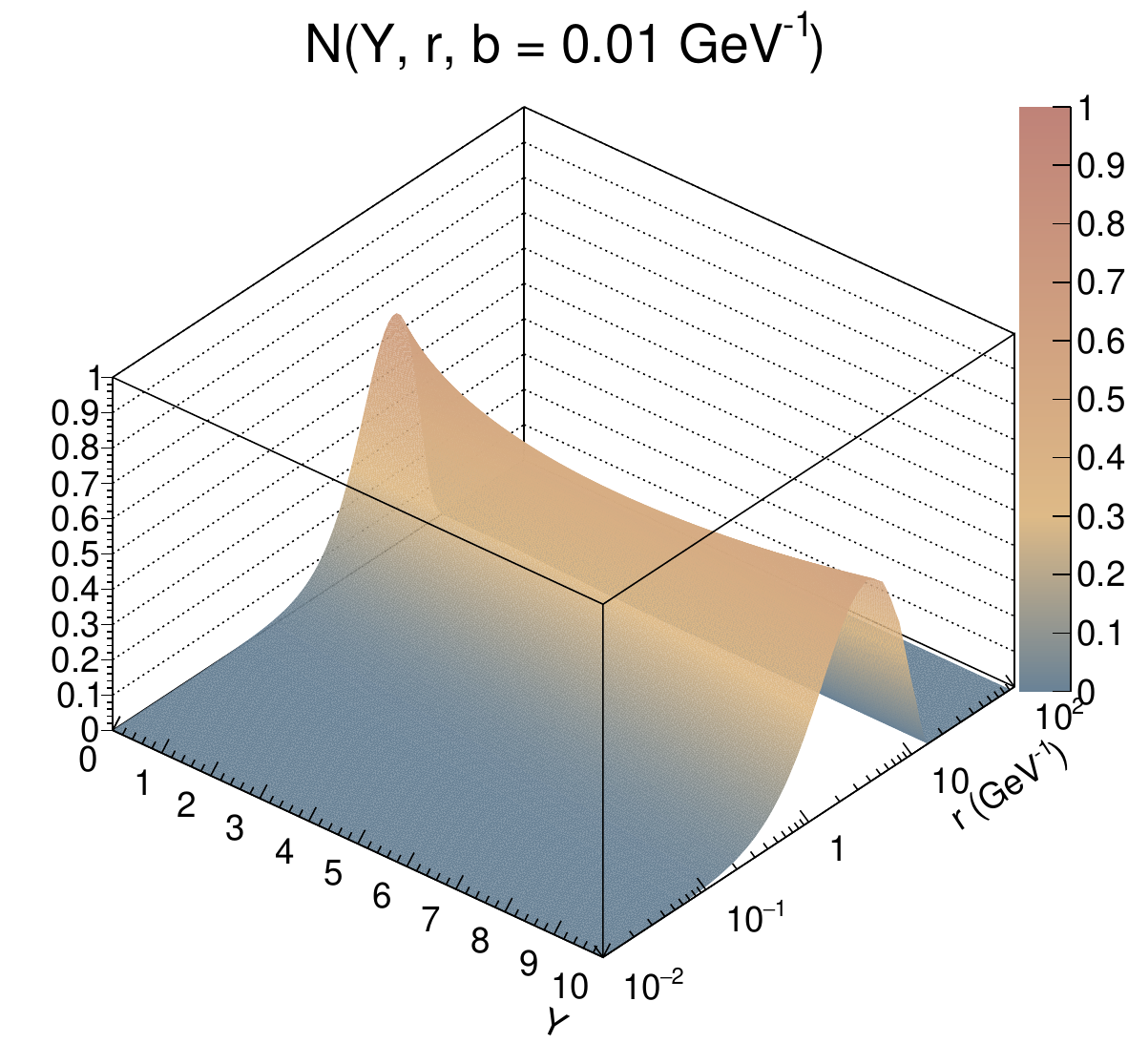} 
    \includegraphics[width=.49\linewidth]{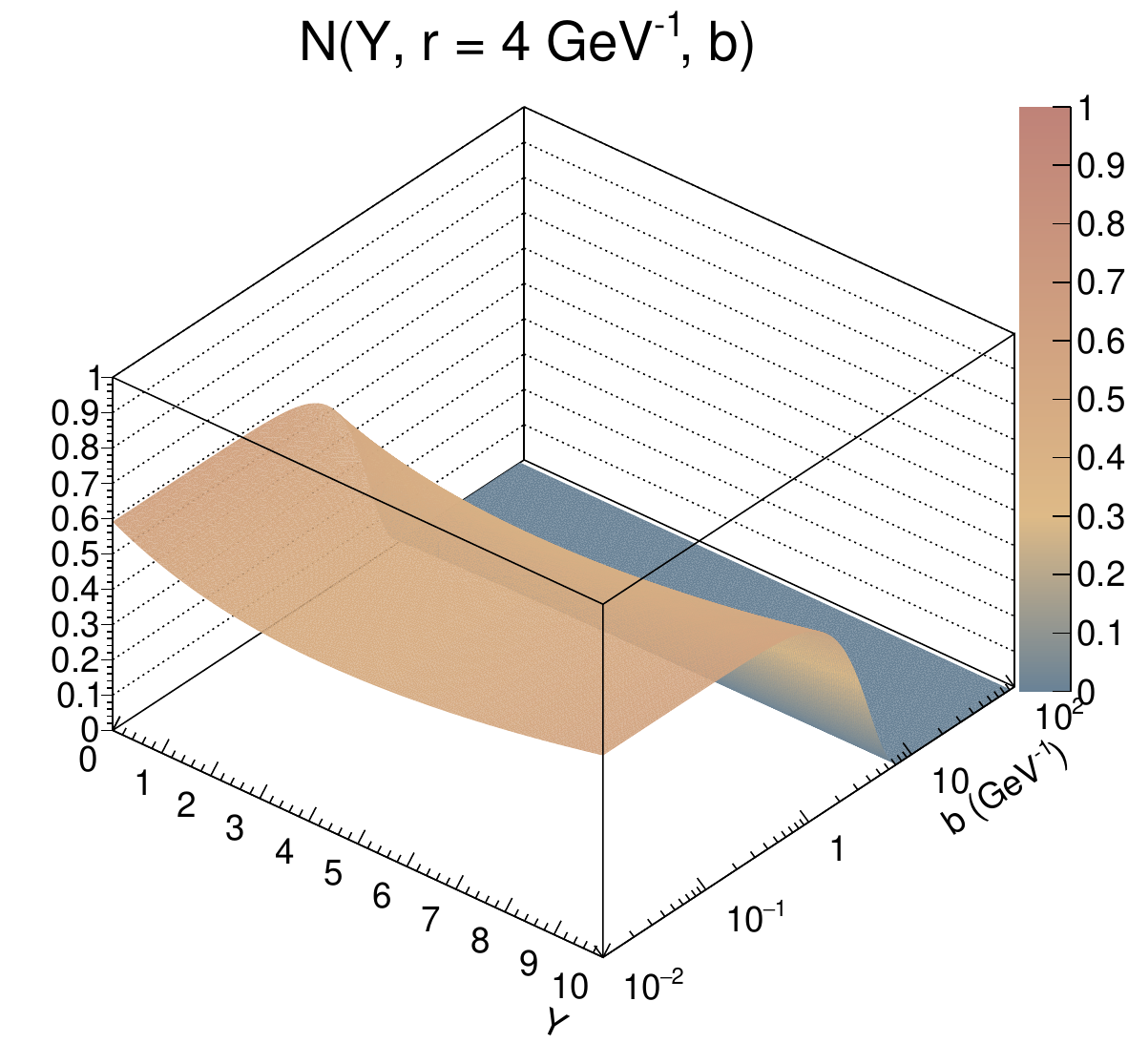}
    \caption{\label{fig:2d}The evolution of dipole scattering amplitude $N(Y, r, b)$ according to the NLO BK equation is shown at two fixed values of rapidity in the top panels: $Y = 0$ (top left), and $Y = 10$ (top right). A detailed view of the rapidity evolution on $r$ and $b$ for fixed representative values of $b=0.01$\,GeV$^{-1}$ and $r=4$\,GeV$^{-1}$ is shown in the bottom left, respectively right, panels.}
\end{figure*}

The evolution of $N(Y, r, b)$ according to the NLO BK equation, from the initial condition $Y=0$ to $Y=10$, is shown in the upper panels of Fig.~\ref{fig:2d}. The evolution is very smooth with the dipole amplitude retaining its shape as the rapidity increases. The evolution in $r$ of the forward wavefront is clearly seen, while the backward front changes very little. The shape as a function of the dipole size shows little dependence on the impact parameter for small and moderate values of $b$, while there is a mild evolution on the support at large impact parameters. 

To have a more detailed picture of the evolution, the lower two panels of Fig.~\ref{fig:2d} show the rapidity dependence of the evolution in $r$ (bottom left) and $b$ (bottom right) for fixed representative values of $b=0.01$\,GeV$^{-1}$ and $r=4$\,GeV$^{-1}$, respectively. In both cases one can observe that the maximum of the amplitude first decreases, and then after a few rapidity units it increases. This behaviour, along with the observation that the value of the dipole scattering amplitude does not reach one, has interesting implications for the definition of the saturation scale discussed in Sec.~\ref{sec:Qs}. 
\begin{figure*}[!ht]
    \includegraphics[width=.49\linewidth]{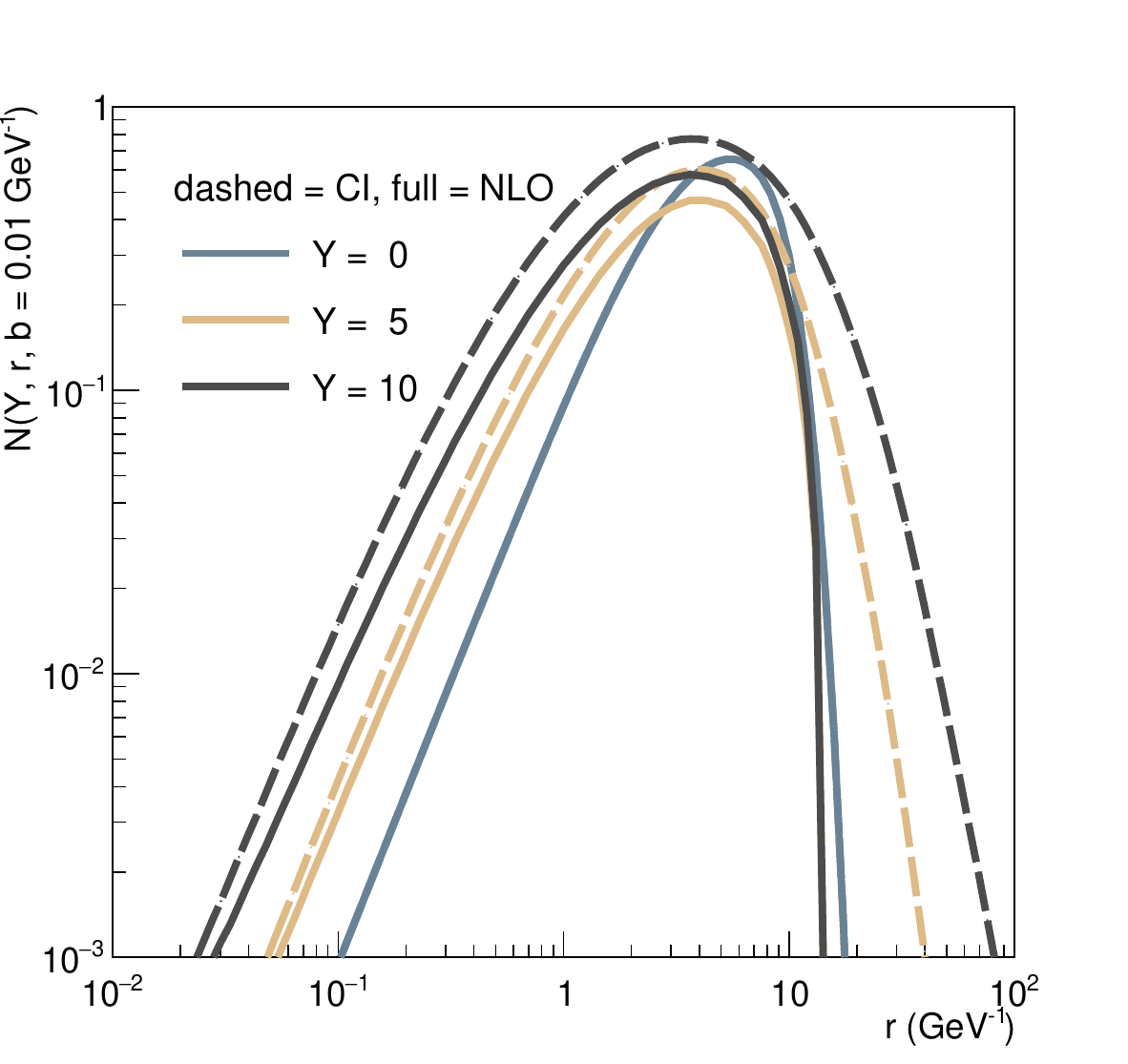}
    \includegraphics[width=.49\linewidth]{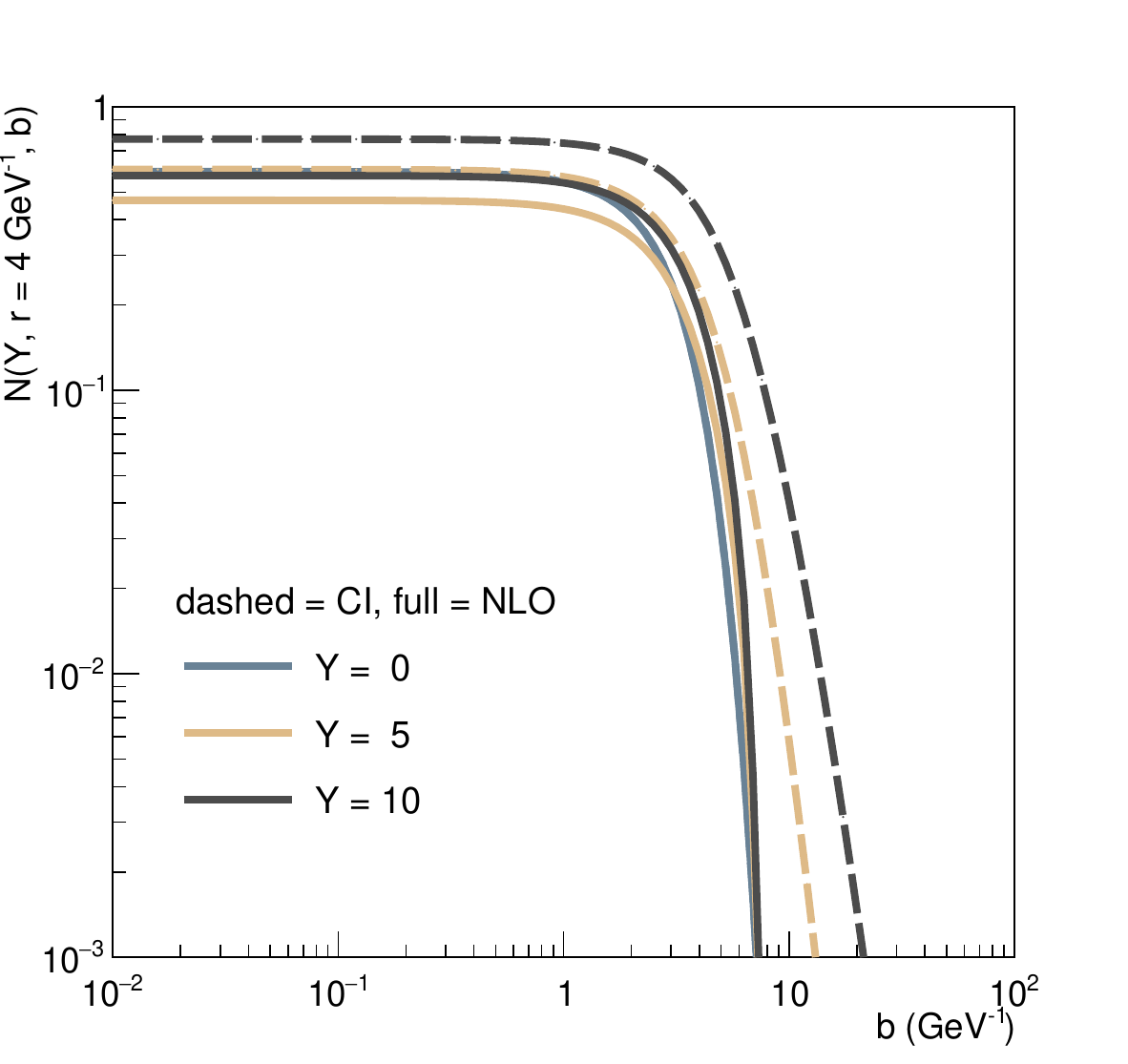}
    \caption{\label{fig:nloXlo_N}Comparison of the dipole scattering amplitude obtained within the collinearly improved (CI) and next-to-leading order (NLO) evolutions for a fixed value of $b = 0.01$\,GeV$^{-1}$ (left) and $r = 4$\,GeV$^{-1}$ (right). Three different values of rapidity are shown.}
\end{figure*}
Figure~\ref{fig:nloXlo_N} shows $r$ and $b$ slices of the scattering dipole amplitudes for the CI and NLO BK equations at three values of rapidity, $Y=0,5,10$, at fixed values of $b=0.01$\,GeV$^{-1}$ and $r=4$\,GeV$^{-1}$, respectively. The behaviour described before is easier to see in this figure: the forward wavefront advances faster than the backward one, and the maximum value of $N$ first decreases and then increases with the behaviour independent of the impact parameter for most of the range. The figure also compares the evolution at NLO with that of the CI equation. At NLO, the evolution is substantially slower than for CI. This behaviour applies to both fronts in $r$ as well as for the decay at large values of the impact parameter. 

An important observation from the right panel of Fig.~\ref{fig:nloXlo_N} is that for the NLO evolution, there are no Coulomb tails. This behaviour is puzzling. We expected that, as in the case of CI evolution, the Coulomb tails would be suppressed with respect to what was found using previous kernels~\cite{Golec-Biernat:2003naj}. For the case of CI solutions, the suppression was explained~\cite{Cepila:2018faq} by the confluence of two contributions: the CI kernel that suppresses large daughter dipoles and the chosen initial condition that does not include Coulomb tails. A similar conclusion was found in Ref.~\cite{Contreras:2019vox}, where it was clarified that this is likely a transient behaviour and the tails would reappear at larger rapidities, as commonly expected, see e.g. Ref.~\cite{Gotsman:2019ehe}. The initial condition chosen for the solution of the NLO BK equation is similar to what was used before, and, as seen above, the NLO evolution is slower, so smaller tails were expected. There, however, seems to be none. This behaviour is further discussed in Section~\ref{sec:nlo_dds}, where a potential explanation is offered.

\subsection{Anomalous dimension and evolution speed}
Here, we present two important characteristics of the evolution of the dipole amplitude: the anomalous dimension and the evolution speed. In both cases, the behaviour of the NLO BK equation solutions is further compared to that of the CI case.

\begin{figure}[!ht]
    \includegraphics[width=0.49\linewidth]{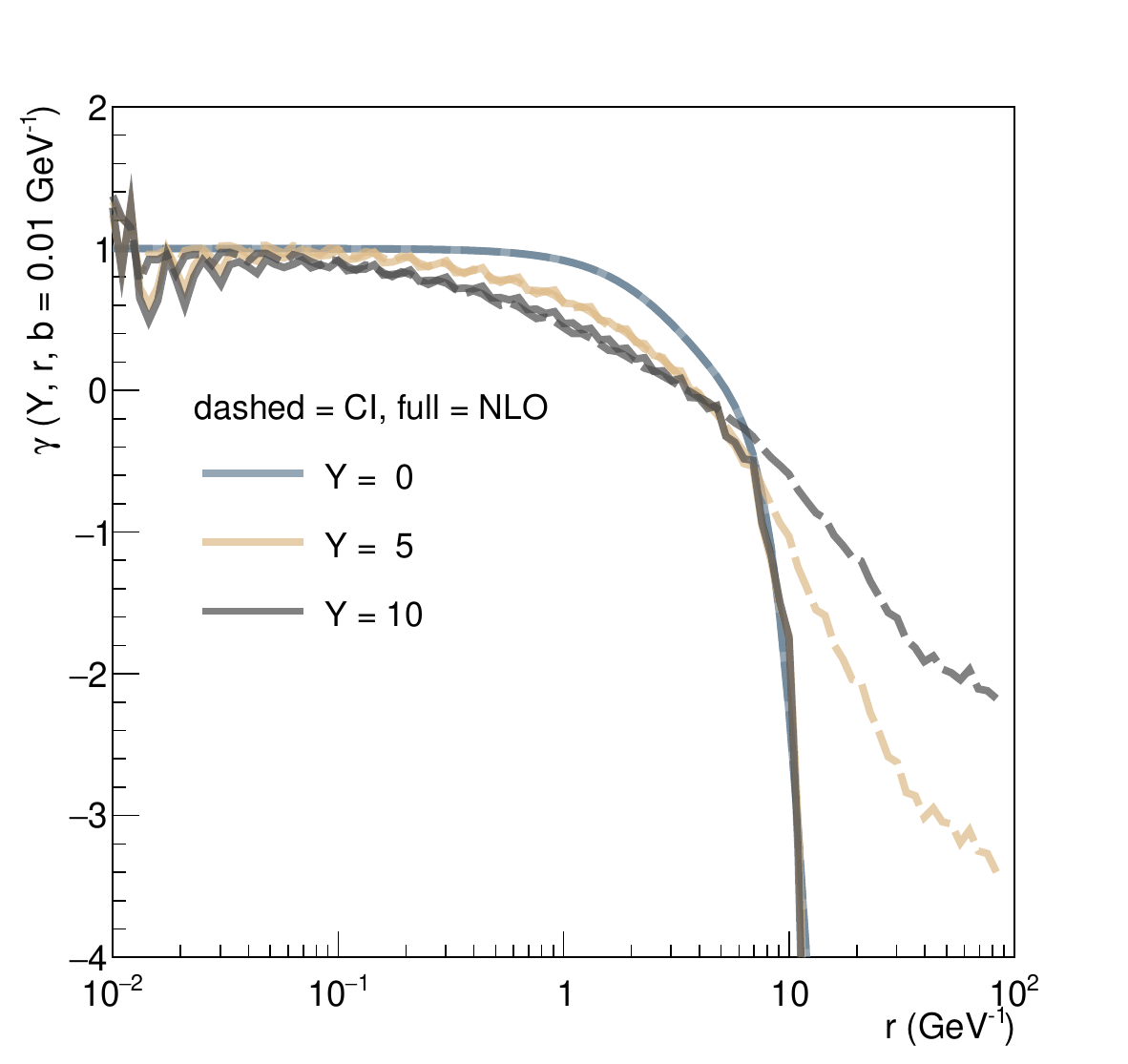}
    \caption{\label{fig:anom_dim}Anomalous dimension $\gamma$ as a function of the dipole size at a representative impact parameter $b = 0.01 \mathrm{~GeV}^{-1}$ and three points of the rapidity evolution. Dashed curves show the leading order results (CI), while the NLO results are plotted in full curves. }
\end{figure}

The anomalous dimension is defined as
\begin{equation}
    \gamma(Y, r, b) = \frac{\partial \ln N (Y, r, b)}{\partial \ln r^2},
\end{equation}
and it is shown in Fig.~\ref{fig:anom_dim} as a function of the dipole size $r$ at a representative value of the impact parameter.
The anomalous dimension is flat for the NLO dipole amplitudes for most of the range in dipole sizes. For large dipoles, the anomalous dimension decreases quite rapidly. Upon evolution, the anomalous dimension seems to slightly decrease its value at medium and small dipole sizes, but the numerical fluctuations are large compared to this effect. The anomalous dimension for the CI case shows the same behaviour. At large values of the dipole size (above 10\,GeV$^{-1}$), the shape of the anomalous dimension is markedly different between both cases. This is related to the fact that for the NLO evolution, there is almost no change for the backward front of the dipole amplitude, as mentioned above.

\begin{figure}[!ht]
    \centering
    \includegraphics[width=.49\linewidth]{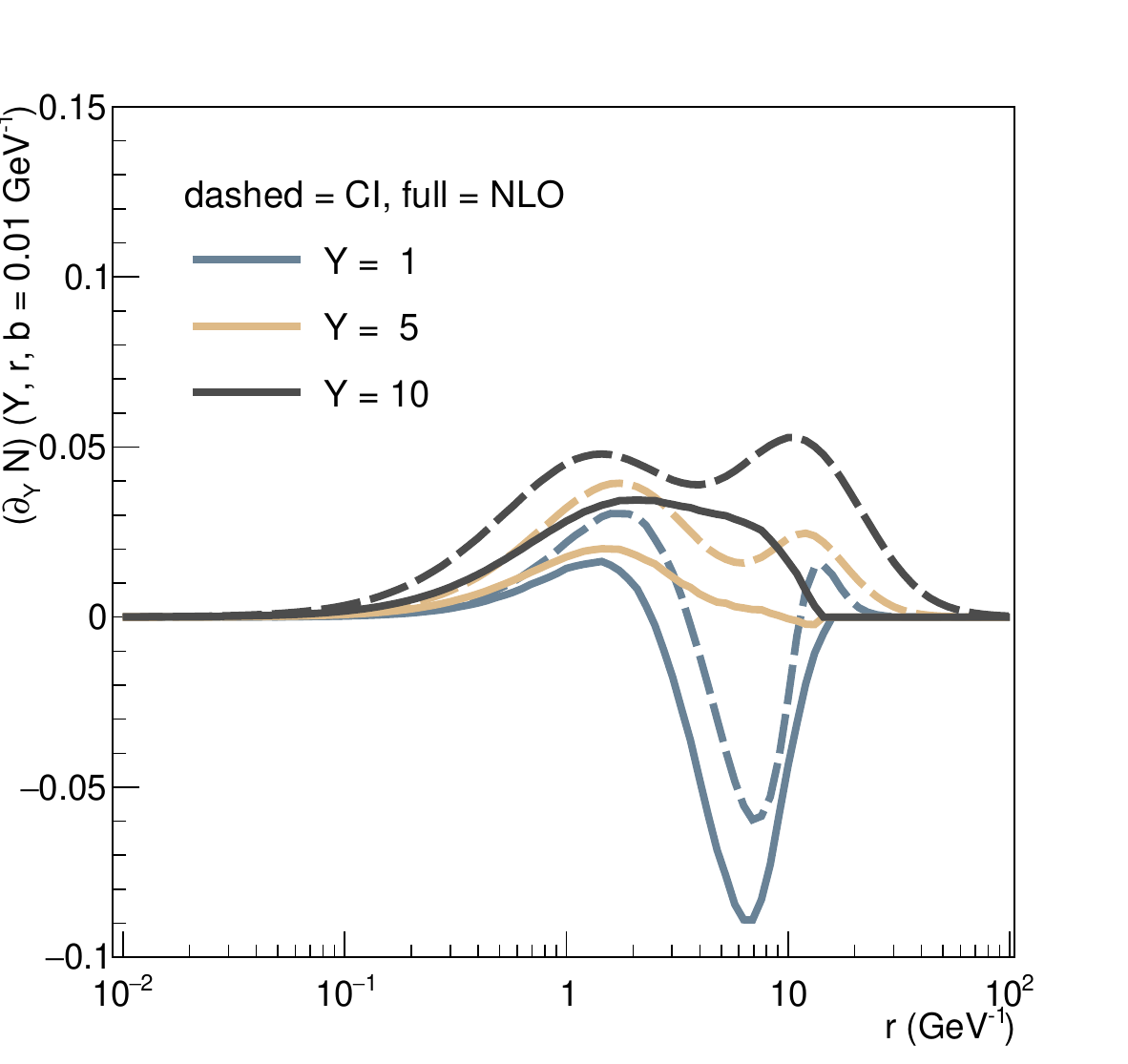}
    \caption{\label{fig:evo_speed}The evolution speed as a function of the dipole size $r$ at a representative  impact parameter $b = 0.01 \mathrm{~GeV}^{-1}$ and three steps of the rapidity evolution. The CI BK corresponds to the dashed curves, the NLO BK is shown in full.}
\end{figure}

\begin{figure*}[!ht]
    \includegraphics[width=.49\linewidth]{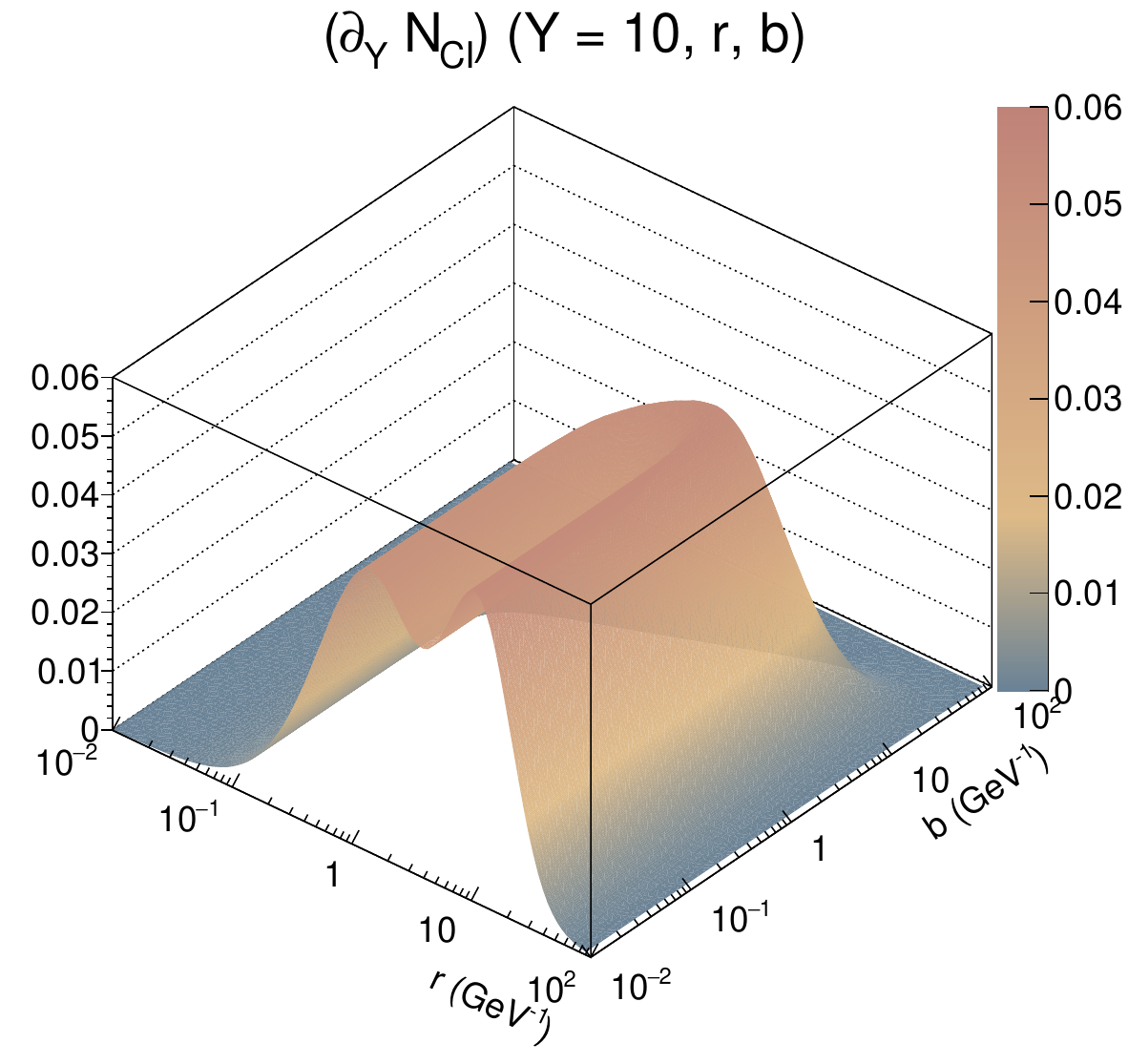}
    \includegraphics[width=.49\linewidth]{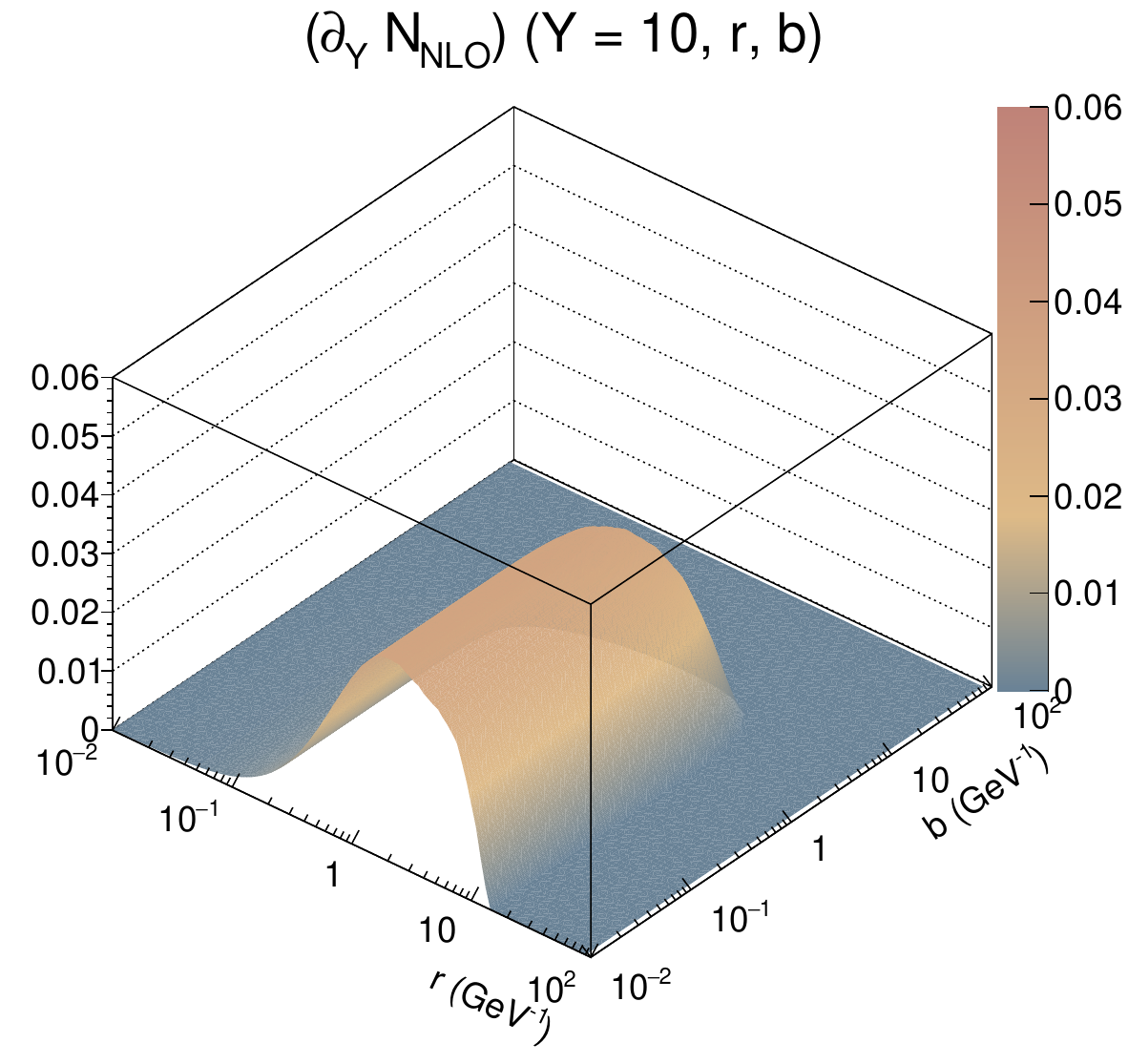}
    \caption{\label{fig:evo_speed2d}The evolution speed as a function of the dipole size $r$ and impact parameter $b$ at rapidity $Y = 10$ for the CI BK (left panel) and the NLO BK (right panel).}
\end{figure*}

Another important characteristic of the solutions to the BK equation is the evolution speed: $\partial_Y N(Y, r, b)$, which is shown in Fig.~\ref{fig:evo_speed} for a fixed impact parameter $b = 0.01 \mathrm{~GeV}^{-1}$ and three different rapidities. The evolution speed for dipole sizes above around 2\,GeV$^{-1}$ is negative and at even larger dipole sizes shows a different behaviour for CI and NLO amplitudes. At sizes below around 2\,GeV$^{-1}$, that is for the forward wavefront, the evolution speed is noticeably slower for the NLO evolution with respect to the CI case. To demonstrate that this behaviour is similar at other impact parameters, we show in Fig.~\ref{fig:evo_speed2d} the evolution speed at rapidity 10 as a function of both the dipole size and the impact parameter for CI (left) and NLO (right) dipole amplitudes.

\subsection{\label{sec:Qs}Saturation scale}
Now, we turn to the saturation scale $Q_\mathrm{s} = Q_\mathrm{s}(Y,b)$ defined as
\begin{equation}
    N(Y, r = 1 / Q_\mathrm{s}, b) = \kappa,
    \label{eq:sat_scale}
\end{equation}
and its dependence on the impact parameter and rapidity. The constant $\kappa$ is of the order of unity.
Two values for $\kappa$ have been traditionally used in the past: 0.5 or $1-\exp{(-1/2)} \approx 0.4$. Looking again at the upper left panel of Fig.~\ref{fig:nloXlo_N} we see that there are rapidities for which the maximum of the amplitude does not reach 0.5. This would produce a situation where the saturation scale is defined at the initial condition, then there is no saturation scale, and finally, at larger rapidities, the saturation would again appear. To avoid this, we decided to use $\kappa=0.4$.

\begin{figure*}[!ht]
    \centering
    \includegraphics[width=.49\linewidth]{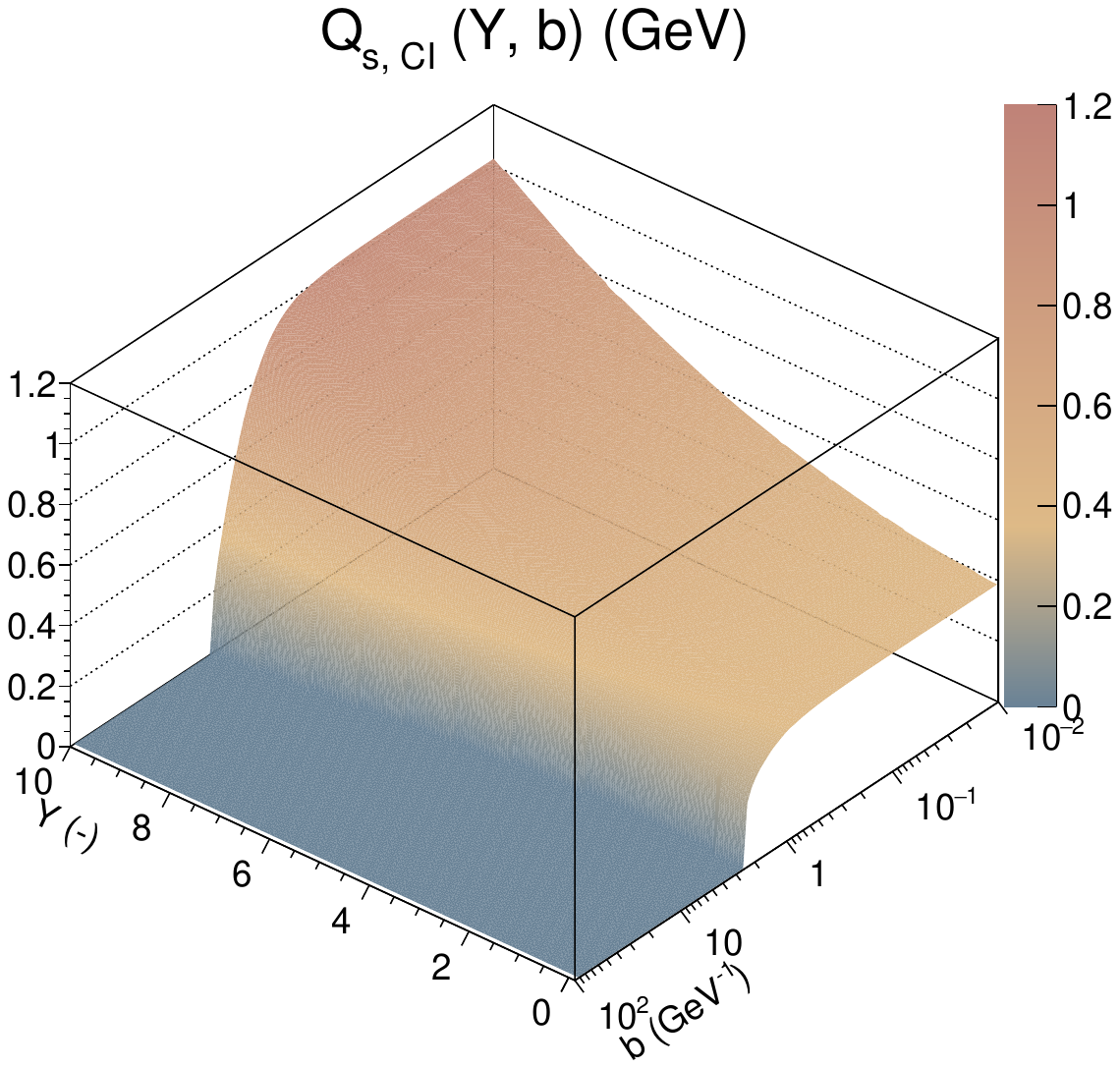}
    \includegraphics[width=.49\linewidth]{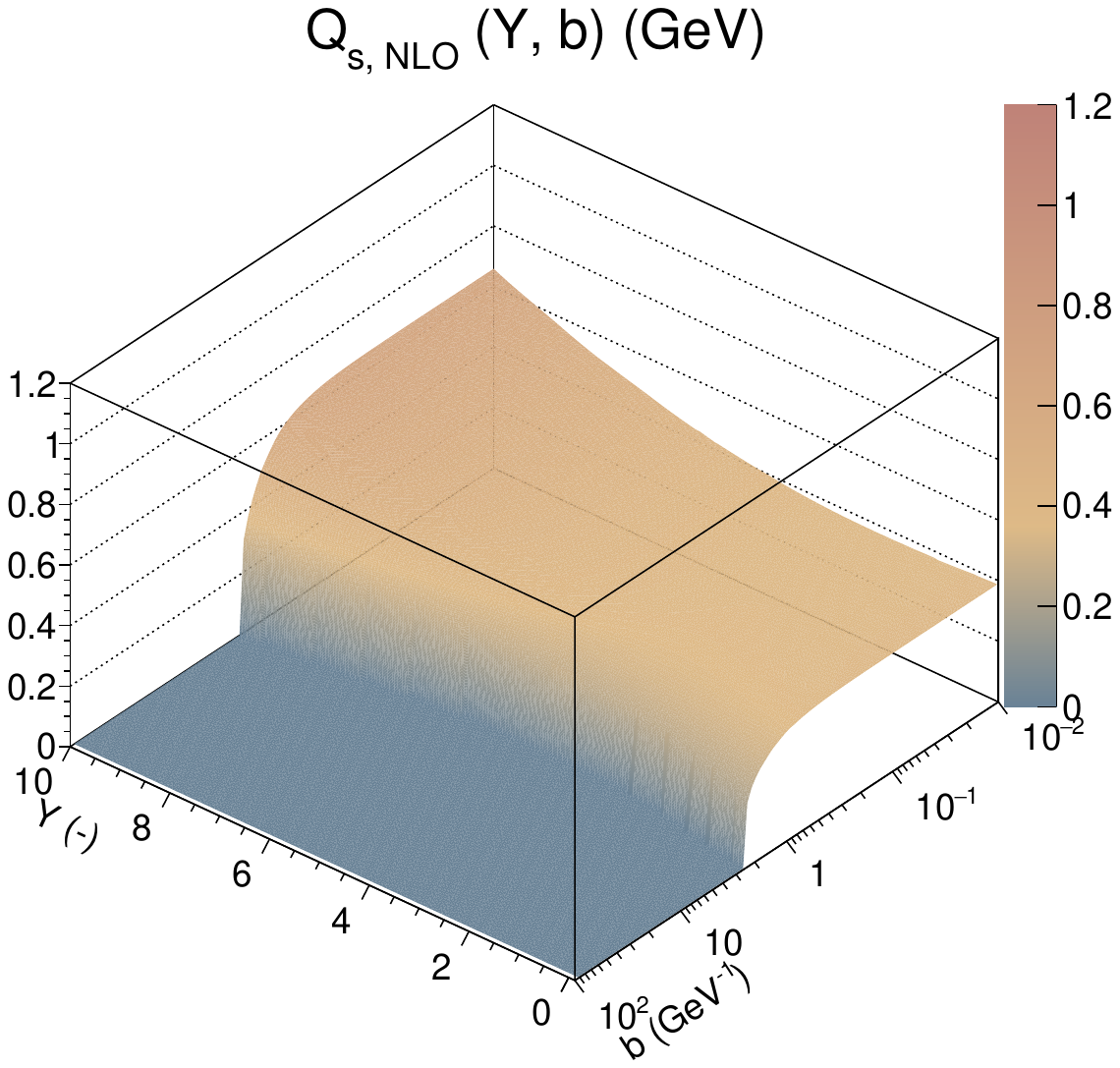}
    \caption{\label{fig:sat_scale}The evolution of the saturation scale for the solutions of the CI (left) and NLO (right) BK equations from $Y=0$ to $Y=10$ is shown as a function of the dipole size and the impact parameter.}
\end{figure*}

The behaviour of the saturation scale in CI and NLO BK is shown in Fig.~\ref{fig:sat_scale}. For impact parameters below around 1\,GeV$^{-1}$, the saturation scale is constant in both cases. The dependence of $Q_\mathrm{s}$ on rapidity is steeper for the CI case, which is expected, given the observations discussed in the previous section. 
For $Y = 10$, the saturation scale at small to moderate impact parameters reaches a value of 0.7 GeV for the NLO case while it is above 1 GeV for CI, after having started in both cases at a value of 0.4 GeV at the initial rapidity.

\subsection{\label{sec:nlo_dds}NLO daughter dipoles}
To address the disappearance of Coulomb tails in our solution and pinpoint the origin of this behaviour, we now discuss the early evolution speed and its partial components. 

\begin{figure}[!ht]
    \centering
    \includegraphics[width=0.49\linewidth]{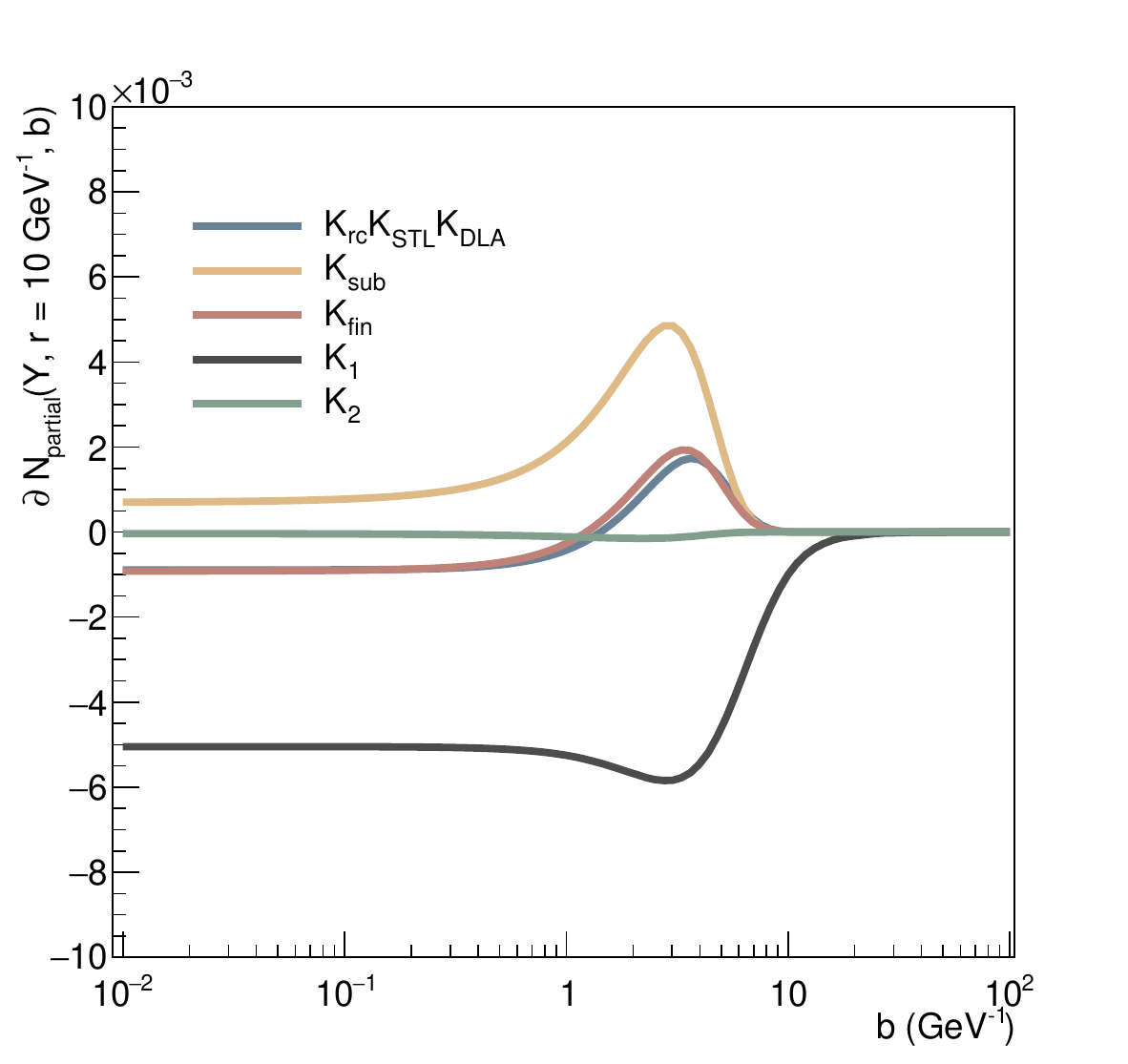}
    \includegraphics[width=0.49\linewidth]{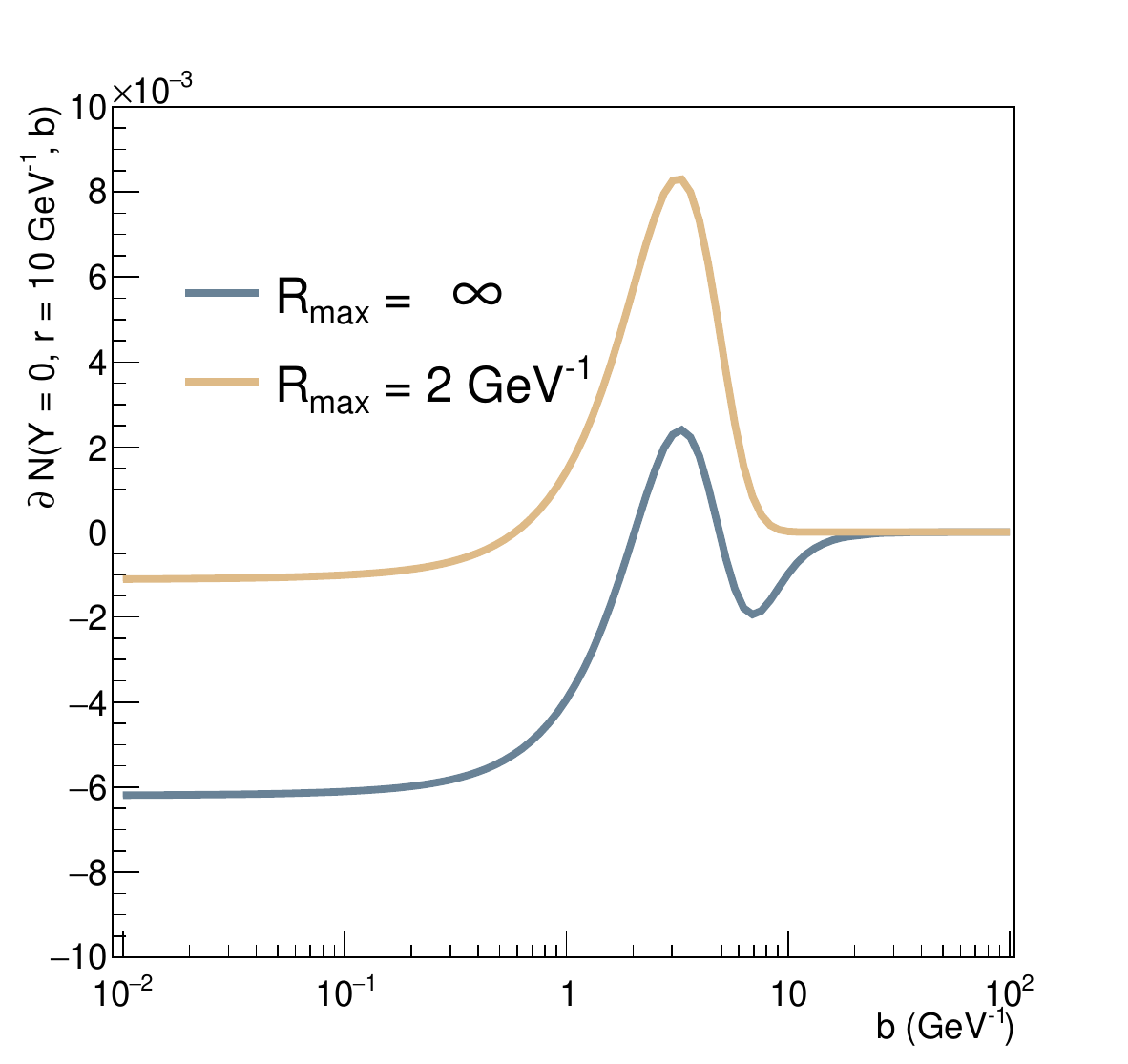}
    \caption{\label{fig:partial}
    Contributions to the difference between the initial condition and the first step in rapidity ($Y = 0.05$) from various parts of the NLO kernel are shown in the left panel. The right panel shows the total contribution for the original set-up (sum of the partial ones in the left panel) and one where the NLO daughter dipoles are cut off.}
\end{figure}

A step of the evolution is given by the rapidity step $\Delta Y$ multiplied by $\partial_{} N(Y)$, the r.h.s. of Eq. (\ref{eq:nlo-bk}), as shown schematically in 
Eq.~(\ref{eq:partial}). 
This can be split into partial contributions as 
\begin{eqnarray}
\label{eq:partial}
    N(Y + \Delta Y) &&= \Delta Y \partial_{}N(Y) 
    \nonumber\\ &&= 
    \Delta Y \big( \partial_{}N_{\rm rc,STL,DLA}(Y) + \partial_{}N_{\rm sub}(Y) + \partial_{}N_{\rm fin}(Y) + \partial_{}N_{1}(Y) + \partial_{}N_{2}(Y)\big).
\end{eqnarray}
to separately study the effects of the corresponding kernels. The individual partial contributions to the first evolution step are shown in the left panel of Fig.~\ref{fig:partial}, labeled by the corresponding part of the kernel. 
The sum of these partial contributions $\partial N$ is shown by the blue curve in the right panel of Fig.~\ref{fig:partial}.
In the region of Coulomb tails, i.e. at $b \approx 10 {\rm~GeV}^{-1}$, this total contribution is negative, leading to suppression of the dipole amplitude. From the left panel of Fig.~\ref{fig:partial} we can then identify the origin of this suppression with the contribution of the $K_1$ kernel.
 
This behaviour leads to obtaining small negative values during the numerical calculation. These are, however, nonphysical and can spoil the stability of the equation, and, as explained in Sec.~\ref{sec:sol}, are therefore set to 0. 
A similar situation occurred in the previous attempts to solve the NLO BK equation~\cite{Lappi:2015fma}. In that case, it was addressed in Ref.~\cite{Lappi:2016fmu} by including the resummation of the collinear contributions, which imposes proper time ordering and thus suppresses large daughter dipoles. 
This, however, only applies to the LO daughter dipoles. For the NLO daughter dipoles (connected to the second gluon emission), a similar approach can be employed by imposing a cut-off in mass $m$ (see e.g. Ref.~\cite{Berger:2011ew}) in the form of $R_{\rm max} = 1/m$ as
\begin{align}
    K_{1} &\rightarrow K_1~\uptheta(R_{\rm max} - r_{xw})\uptheta(R_{\rm max} - r_{wy})\uptheta(R_{\rm max} - r_{zw}),
    \\
    K_{2} &\rightarrow K_2~\uptheta(R_{\rm max} - r_{xw})\uptheta(R_{\rm max} - r_{wy})\uptheta(R_{\rm max} - r_{zw}).
\end{align}
Whenever an NLO daughter dipole is larger than the cut-off value $R_{\rm max}$, the Heaviside function $\uptheta (r)$ suppresses the contribution from the corresponding part of the kernel.

To describe the qualitative impact of this cut-off, a range of $R_{\rm max}$ values was tested, from those with no observable effect to those suppressing the NLO kernels too strongly. 
In the right panel of Fig.~\ref{fig:partial}, we show how the total evolution contribution is modified, in particular in the region of Coulomb tails, for $R_{\rm max} = 2 {\rm~GeV}^{-1}$.

\begin{figure}[!ht]
    \centering
    \includegraphics[width=0.49\linewidth]{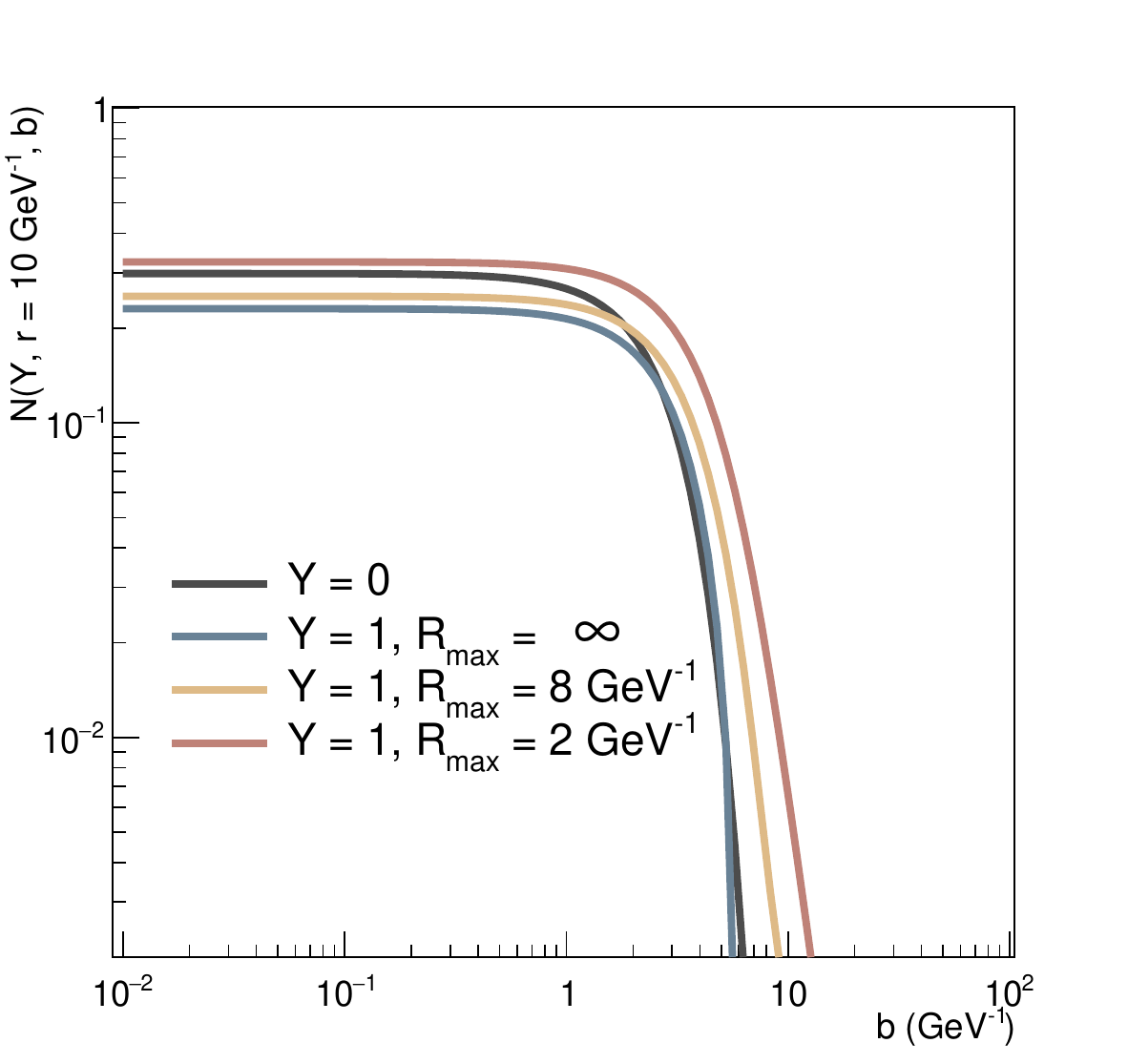}
    \caption{\label{fig:coulomb_tails}
    A comparison of the evolution to rapidity $Y=1$ with and without cutting off the large NLO daughter dipoles. Unlike the blue curve without the cut-off, the gold and red curves grow the Coulomb tails based on the value of the $R_{\rm max}$ cut-off.
    }
\end{figure}

In Fig.~\ref{fig:coulomb_tails} we show the evolution of the initial condition to rapidity $Y = 1$ for \mbox{$R_{\rm max} = 2 {\rm~GeV}^{-1}$} and $R_{\rm max} = 8 {\rm~GeV}^{-1}$. For comparison, the evolution without cut-off, effectively corresponding to $R_{\rm max} \rightarrow \infty$, is also shown. In the last case, the Coulomb tails are removed completely. Note that using $R_{\rm max} = 2 {\rm~GeV}^{-1}$ has the added benefit that the number of cases where the dipole amplitude takes slightly negative values is strongly reduced.

We see that restricting the contributions from the large NLO daughter dipoles in the evolution can be used to reintroduce the contribution at large impact parameters to the dipole amplitude, while still suppressing the Coulomb tails as in the CI case~\cite{Cepila:2018faq}. The value of the cut-off parameter $R_{\rm max}$ governs quantitatively this behaviour. A side effect of this modification is a change in the normalisation of the total amplitude, as can also be seen in Fig.~\ref{fig:coulomb_tails}.

We have shown that cutting off the contribution of the large NLO daughter dipoles has an essential impact on the behaviour of the dipole amplitude at large impact parameters. The use of a cut-off, although effective, is rather crude. This suggests the importance of finding a proper way to analytically re-sum the collinear contributions from the NLO daughter dipoles, as was already done for the LO daughter dipoles, and find the corresponding modification of the kernel of the NLO BK equation.

\section{Summary}
This paper presents the first stable solution of the next-to-leading order BK equation, including the dependence on the dipole size as well as on the impact parameter. We described the numerical method allowing for a stable solution, namely the proper choice of shifted integration grids. This way, the contribution of divergent kernel points can be minimised, and the only necessary numerical intervention is to hard restrict the dipole amplitude values $N(Y, r, b) \in [0,1]$.

The solution to the leading-order equation BK equation with collinearly improved kernel (CI) was shown to compare with the NLO evolution. Overall, the NLO evolution is slower and suppresses the contribution of both large dipoles and large impact parameters. The almost absolute suppression of the Coulomb tails is, however, stronger than expected and encourages further efforts in resumming the collinear contributions from the NLO daughter dipoles.

To characterise the amplitude evolution, the anomalous dimension and evolution speed are plotted and discussed from the perspective of both the NLO and CI BK equations. The evolution speed of the NLO BK is overall smaller than in the CI case, and as the rapidity increases, the NLO anomalous dimension is shown to be more stable.

Finally, the evolution of the saturation scale is compared again for the NLO and CI BK equation, quantifying the different regions in which the saturation effects are considered to take effect.

These results present a breakthrough in numerical solutions of the BK equation, paving the way to its application to compute observables to search for saturation with next-to-leading order precision.

{\bf Data availability:}
The scattering amplitudes are publicly available at the CERN-based open repository ZENODO \cite{vaculciak_2024_14362815}.

\begin{acknowledgements}
We thank Heikki Mäntysaari for early discussions about the contribution of large NLO daughter dipoles, and the anonymous referee for useful suggestions to explore the suppression of the Coulomb tails.
This work was partially funded by the Czech Science Foundation (GAČR), project No. 22-27262S. Marek Matas was furthermore supported by the CTU Mobility Project MSCA-F-CZ-III under the number \texttt{CZ.02.01.01/00/22\_010/0008601}.
\end{acknowledgements}

\appendix 
\section{Numerical solution of the BK equation at next-to-leading order}\label{app:solving}
Here we summarize the numerical approach to solving the NLO BK equation. Using Simpson's integration method enabled good tracking of points, causing the numerical instabilities in the BK equation, allowing us to pinpoint the main problem to the overly symmetric sampling of the integration space. 
\begin{figure}[!ht]
    \includegraphics[width=0.49\linewidth]{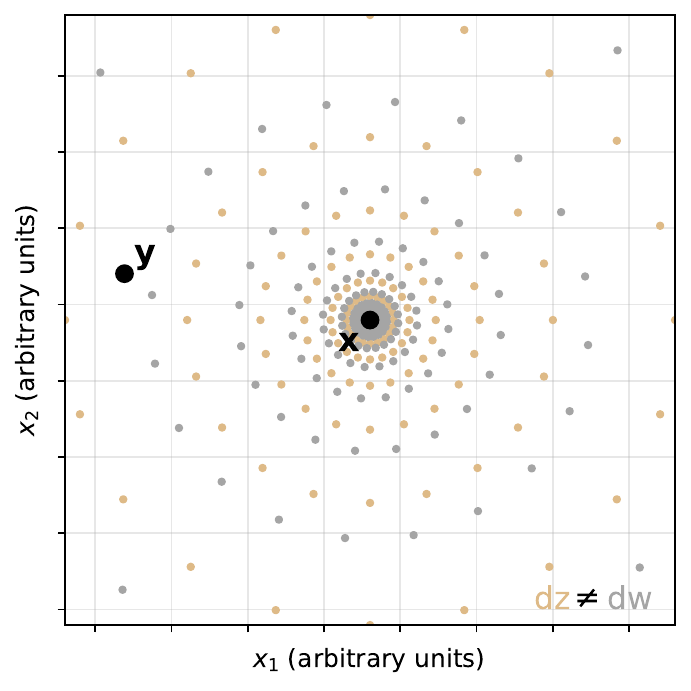}
    \caption{\label{fig:grids}Schematic structure of the integration grids in the transverse plane, same as in Fig.~\ref{fig:layout}. Both grids (in $dz$ and $dw$) are polar with logarithmic spacing in the radial coordinate, centred around one of the quarks. To ensure uniform sampling of the integrand with less emphasis on the high-symmetry divergent points, they are offset w.r.t. each other and the sampling grid. 
    The grids in this picture have been adjusted for better clarity and do not have the actual number of points.}
\end{figure}

\begin{figure*}
    \centering
    \includegraphics[width=0.49\linewidth]{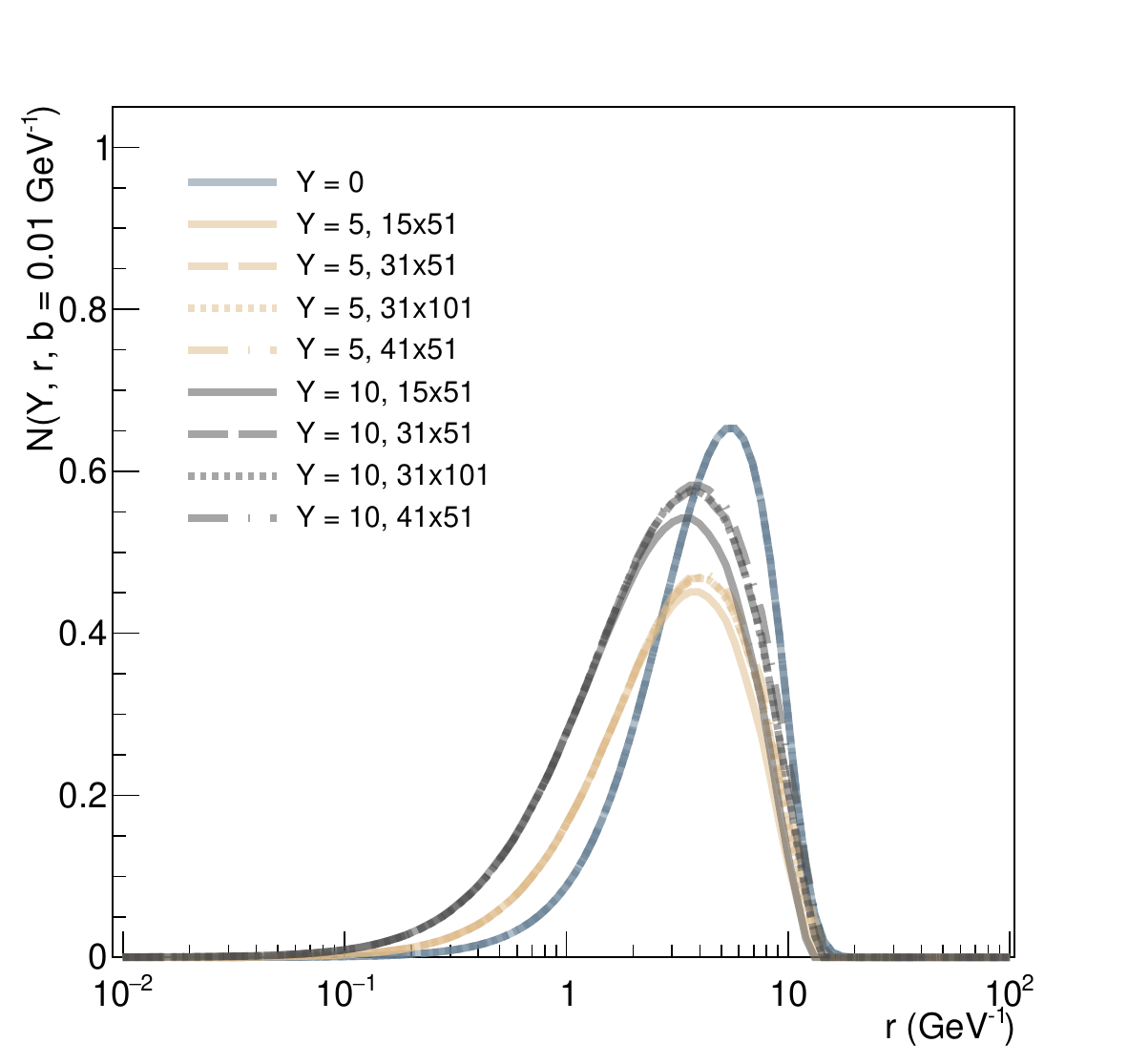}
    \includegraphics[width=0.49\linewidth]{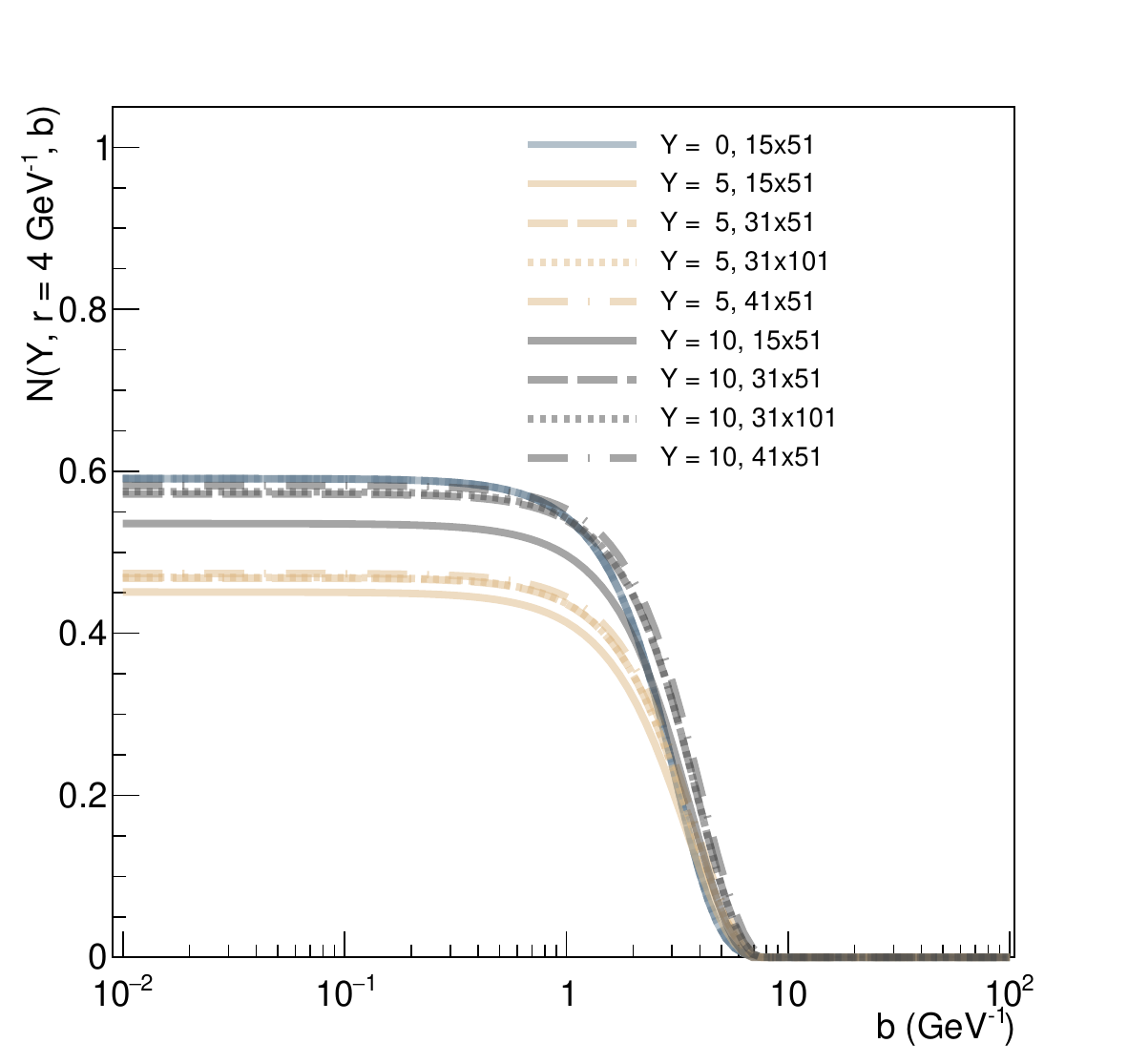}
    \caption{\label{fig:hyperpar}The dipole amplitude at various stages of the evolution using $n_a \times n_r$ points in integration grids, where $n_a$ is the number of angular points and $n_r$ is the number of radial points. Clearly, the $31 \times 51$ set-up is fine enough for a convergent solution.}
\end{figure*}

The key ingredient for overcoming the instability was therefore identified as displacing the sampling (the points in the transverse plane in which the quark and antiquark of the mother dipole are placed during the calculation) and both integration (the points in the transverse plane in which the emerging gluons, vertices of the daughter dipoles, are placed during the calculation) grids as shown in Fig.~\ref{fig:grids}. This way, quarks and gluons not only never emerge at the same point (giving zero dipole sizes), but rarely hit a symmetrical configuration, in which e.g. $r_{xz}^2 r_{wy}^2 - r_{xw}^2 r_{zy}^2 = 0$. Omitting these (and adjacent) configurations, in which the kernels diverge (analytically or numerically because of the limited numerical precision) gives a solid numerical set-up to solve the NLO BK equation. Furthermore, Fig.~\ref{fig:hyperpar} shows the stability of the solution by varying the computational set-up.

The integration grids are polar with linear sampling in the angular variable $\varphi$ and logarithmic in the radial $r$, centred at the quark position $x$ (the left quark in Fig.~\ref{fig:layout}.). The displacement then corresponds to shifting the initial and final grid point somewhere close to the middle of the other grid (a place somewhere close to the middle is chosen by dividing the interval by $\sqrt{3}$.), such that the divergence-causing symmetry remains unrecovered and equidistantly, in both the linear and logarithmic sense, sampling the resulting interval, as depicted in Fig.~\ref{fig:grids}. The shifting values were chosen heuristically and are listed in Tab.~\ref{tab:simpson_displacement}.

\begin{table}[!ht]
\centering
\begin{tabular}{ll}
\hline
\textbf{sampling grid} & \textbf{value, sampling} \\
\hline
rapidity $Y$             & lin, [0,10], 201 points \\ 
dipole size $r$             & log, [$10^{-2}$, $10^{2}$], $101$ points \\
impact parameter $b$        & log, [$10^{-2}$, $10^{2}$], $101$ points \\
\hline
\textbf{integration grid (z)} & \textbf{value, sampling} \\
\hline
radius $r_z$                & log, [$1.1 \cdot 10^{-2}$, $1.1 \cdot 10^{2}$], $51$ points \\
angle $\varphi_z$           & lin, [$0.100$, $6.383$], $31$ points \\
\textbf{integration grid (w)} & \textbf{value, sampling} \\
\hline
radius $r_w$                & log, [$1.21 \cdot 10^{-2}$, $1.21 \cdot 10^{2}$], $51$ points \\
angle $\varphi_w$           & lin, [$0.217$, $6.500$], $31$ points \\
\hline
\end{tabular}
\caption{\label{tab:simpson_displacement}Values of the parameters used in the convergent Simpson-based calculation. Values of radius and impact parameter are in the units of GeV$^{-1}$, angular variables are in rad, rapidity and number of samples are dimensionless.}
\end{table}

\bibliography{bibliography}
\end{document}